\documentclass[aps, pra, twocolumn, superscriptaddress, amsmath, tightenlines, longbibliography]{revtex4-2}

\usepackage{amssymb}
\usepackage{amsmath}
\usepackage{dcolumn}
\usepackage{graphicx}
\usepackage{mathrsfs}
\usepackage{appendix}
\usepackage{graphicx}
\usepackage{booktabs}
\usepackage{colortbl}
\usepackage{float}

\definecolor{Dred}{RGB}{190,0,0}

\usepackage{url}
\usepackage[colorlinks]{hyperref}
\hypersetup{%
	plainpages=true,
	breaklinks=true,       %not default in dvips mode, so we must specify
	hypertexnames=false,  %not ideal, but needed when pagenums duplicate (`i' vs. `1')
	pageanchor=true,
	colorlinks=true,
	linkcolor={blue},
	citecolor={red},
	urlcolor={blue},
	anchorcolor={black}
}

\def \hide#1{}

\begin{document}
\title{Circuit QED with Giant Atoms Coupling to Left-handed Superlattice Metamaterials}

\author{Zhao-Min Gao}
\affiliation{Institute of Theoretical Physics, School of Physics, Xi'an Jiaotong University, Xi'an 710049, People’s Republic of China}

\author{Jia-Qi Li}
\affiliation{Institute of Theoretical Physics, School of Physics, Xi'an Jiaotong University, Xi'an 710049, People’s Republic of China}

\author{Zi-Wen Li}
\affiliation{Institute of Theoretical Physics, School of Physics, Xi'an Jiaotong University, Xi'an 710049, People’s Republic of China}

\author{Wen-Xiao Liu}
\affiliation{Institute of Theoretical Physics, School of Physics, Xi'an Jiaotong University, Xi'an 710049, People’s Republic of China}
\affiliation{Department of Electronic Engineering, North China University of Water Resources and Electric Power, Zhengzhou 450046, People’s Republic of China}

\author{Xin Wang}
\email{wangxin.phy@xjtu.edu.cn}
\affiliation{Institute of Theoretical Physics, School of Physics, Xi'an Jiaotong University, Xi'an 710049, People’s Republic of China}

\date{\today}

\begin{abstract} 
Giant atoms, where the dipole approximation ceases to be valid, allow us to observe unconventional quantum optical phenomena arising from interference and time-delay effects. Most previous studies consider giant atoms coupling to conventional materials with right-handed dispersion. In this study, we first investigate the quantum dynamics of a giant atom interacting with left-handed superlattice metamaterials. Different from those right-handed counterparts, the left-handed superlattices exhibit an asymmetric band gap generated by anomalous dispersive bands and Bragg scattering bands. First, by assuming that the giant atom is in resonance with the continuous dispersive energy band, spontaneous emission will undergo periodic enhancement or suppression due to the interference effect. At the resonant position, there is a significant discrepancy in the spontaneous decay rates between the upper and lower bands, which arises from the differences in group velocity. Second, we explore the non-Markovian dynamics of the giant atom by considering the emitter’s frequency outside the energy band, where bound states will be induced by the interference between two coupling points. By employing both analytical and numerical methods, we demonstrate that the steady atomic population will be periodically modulated, driven by variations in the size of the giant atom. The presence of asymmetric band edges leads to diverse interference dynamics. Finally, we consider the case of two identical emitters coupling to the waveguide and find that the energy within the two emitters undergoes exchange through the mechanism of Rabi oscillations.
\end{abstract}
\maketitle
%%%%%%%%%%%%%%%%%%%%%%%%%%  body  %%%%%%%%%%%%%%%%%%%%%%%%%%
\section{Introduction}
In recent years, there has been considerable research interest in the study of giant atoms due to their ability to produce peculiar phenomena in quantum optics. Unlike small atoms, which are typically treated as point-like particles, the size of giant atom is much larger than or comparable to the wavelength of the propagating field, indicating that the dipole approximation is not valid. \cite{RevModPhys.95.015002,RN1,PhysRevA.104.033710,PhysRevA.106.063717,RN229,chen2023giantatom,PhysRevA.106.063703,phy.2022.1054299}. Under these conditions, it becomes essential to consider the phase accumulation between different coupling points \cite{RN232,PhysRevLett.128.223602,Yang_2021}, which leads to a variety of intriguing phenomena, such as frequency-dependent couplings \cite{PhysRevA.90.013837,PhysRevA.104.023712,RN234,PhysRevResearch.4.023198}, decoherence-free interactions \cite{PhysRevLett.120.140404,PhysRevA.107.023705,PhysRevResearch.2.043184,PhysRevA.107.013710}, unconventional bound states \cite{PhysRevA.107.023716,Xiao_2022,PhysRevLett.126.043602,PhysRevResearch.2.043014,PhysRevA.102.033706,PhysRevA.101.053855,jia2023atomphoton} and chiral quantum optics\cite{Wang_2022-1,PhysRevA.105.023712,Chen_2022,du2023decay,PhysRevX.13.021039}. In experimental setups, giant atoms are typically realized in circuit quantum electrodynamics (circuit-QED) platforms ~\cite{PhysRevA.103.023710,Sun_2023,RN124,Zheng_2023,PhysRevA.104.013720,Andersson_2019,gu2023correlated}.

The interaction between giant atoms and conventional waveguides has been extensively explored in previous studies (e.g., see \cite{PhysRevA.90.043817,RN159,wendin2005,RevModPhys.93.025005,RN2,RN3,RN4,PhysRevApplied.11.054062,PhysRevX.11.041043}). In addition to conventional waveguides and cavities, microwave photons can also exist in artificial environments. An emblematic example is circuit-QED metamaterials, where the dispersion properties and vacuum eigenmodes can be freely tailored in experiments. The structured spectra and asymmetric band gaps can be realized in such metamaterials, providing an intriguing platform for exploring QED phenomena with no analog in traditional circuit-QED setups ~\cite{RevModPhys.86.1093,PhysRevE.71.036609,PhysRevB.106.174304,FanAverittPadilla,PhysRevA.100.053853}. For instance, by spatiotemporally modulating the effective impedance, a superconducting quantum interference device metamaterial can be designed as a chiral quantum waveguide ~\cite{Wang_2022}. When combined with transmission lines, we can achieve multimode strong coupling in circuit QED ~\cite{PhysRevLett.111.163601}.

 In conventional band-gap environment, such as photonic crystals \cite{PhysRevApplied.20.L011001,PhysRevX.6.031017} and the Su-Schrieffer-Heeger model \cite{PhysRevB.98.094307,PhysRevB.100.075437}, the two bands $E_{\pm}(k)$ are induced by the same mechanism. Therefore, the group velocities and band curvatures of their two bands are symmetric with respect to the band gap, i.e., $ v_g^+(k)=-v_g^-(k) $ and $\alpha_+(k) = -\alpha_-(k)$ [with $ \alpha_{\pm}(k)=\partial ^2\omega _{\pm}\left( k \right) /\partial ^2k$]. The left-handed superlattice metamaterial (LHSM) in circuit QED exhibits a unique negative index of refraction \cite{Jung_2014,PhysRevA.99.032325,PhysRevLett.111.163601,PhysRevApplied.14.064033,RN233}. This distinctive property arises from the unconventional interchange of capacitance and inductance, distinguishing LHSM from right-handed materials \cite{10.1063/5.0044103,1687908,PhysRevB.74.113105}. When the impedance of the LHSM is modulated periodically, an asymmetric band gap emerges due to the different physical mechanisms: The upper band at $\omega_+(k=\pi)$ is the infrared cutoff  of the anomalous dispersion \cite{PhysRevApplied.11.054062,PhysRevApplied.14.064033}, while the lower band is a result of Bragg scattering induced by periodic impedance modulation.  Because these two bands stem from different mechanisms, their mode properties (for example, the group velocity and band curvature) are asymmetric with $ v_g^+(k) \ne -v_g^-(k) $ and $\alpha_+(k) \ne -\alpha_-(k)$. These unique spectral features may allow to observe unusual dynamics phenomena of giant emitters \cite{RN235,PhysRevLett.103.043902,RN236}.

In this paper, we find several intriguing phenomena in the circuit QED system composed of giant atoms and LHSM. Firstly, we derive the dispersion relation of LHSM and explain the mechanism behind the band gap generated by the left-handed dispersion band and a band caused by Bragg scattering. By considering a transmon coupled to the proposed LHSM waveguide, we derive the Hamiltonian of the system. When assuming that the emitter is resonant with the upper (lower) band, spontaneous emission is enhanced and suppressed periodically due to the interference effect. Given that the emitter’s frequency is outside the continuous dispersion band, an atom-photon bound state forms at each coupling point.  Due to the asymmetric band edges, the interference dynamics inside the two continuous dispersion energy bands exhibit significant differences, with the atomic steady population in the upper band being much larger than that in the lower band. Lastly, we consider the case of two giant atoms and explore how the dipole-dipole interaction can be modulated by the interference effect.

\section{Left-handed Superlattice Metamaterial}
\label{section:1}

\begin{figure}[ht!]
	\centering\includegraphics[width=9cm]{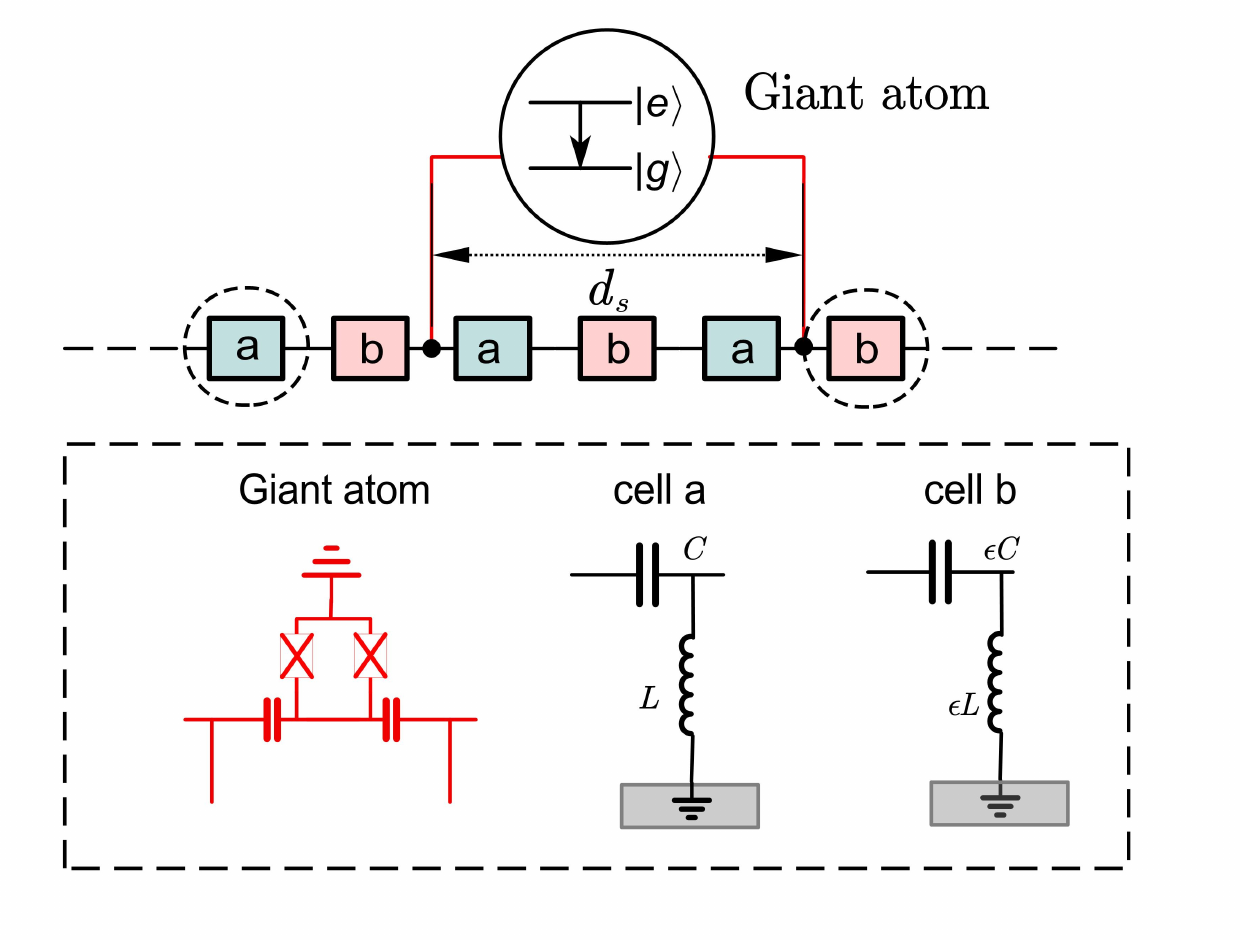}
	\caption{The sketch of a superconducting giant atom coupled to the left-handed superlattice metamaterial. The superlattice cell is composed of two substructures with differing capacitance  $\textit{C}$ ($ \epsilon\textit{C}$) and inductance $\textit{L}$ ($ \epsilon \textit{L}$), represented by cell \rm{a} (\rm{b}).}
	\label{fig1}
\end{figure}

\begin{figure}[hb!]
	\centering\includegraphics[width=7cm]{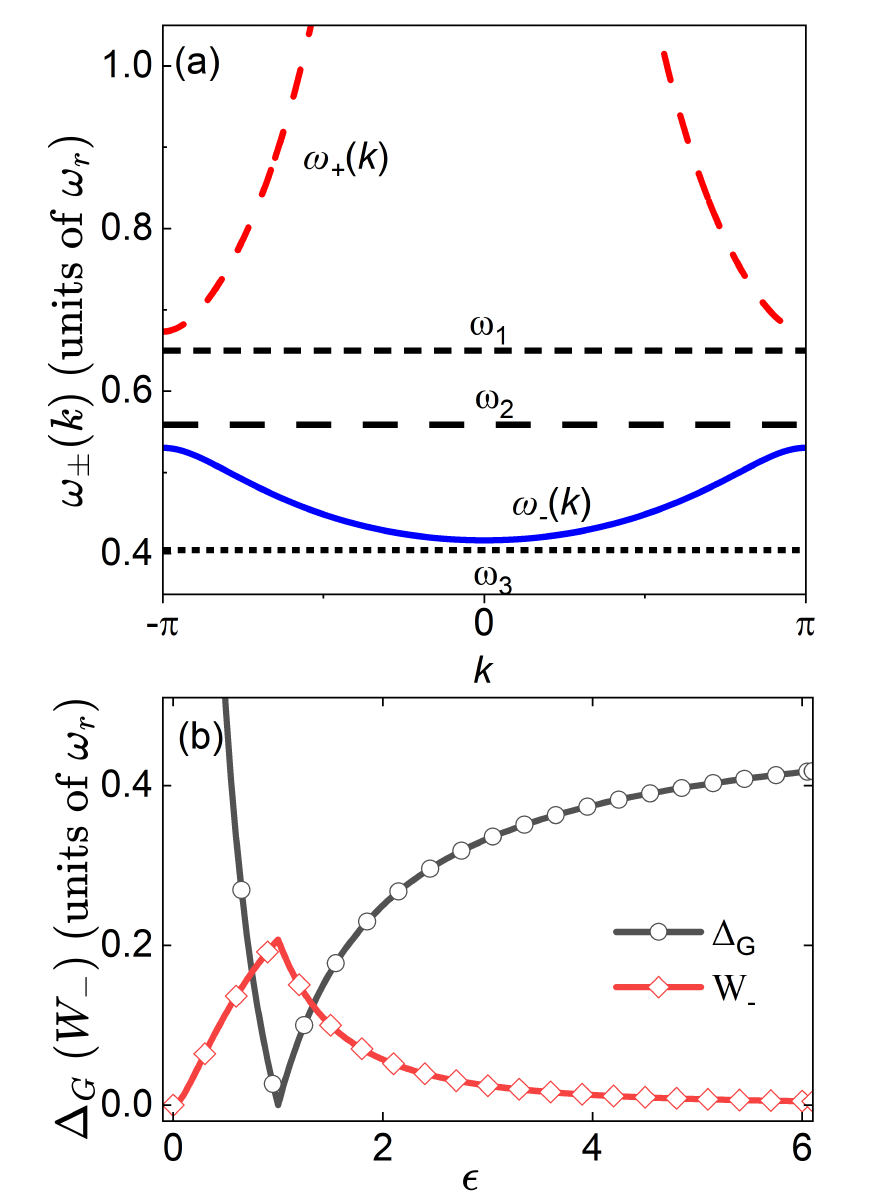}
	\caption{(a) Dispersion relations for two energy bands of the left-handed superlattice metamaterials with $ \epsilon=1.4 $. (b) The width of the lower band, $ W_- $, and the band gap $ \Delta_G $, as functions of the superlattice parameter $ \epsilon $. Parameters of the system are $ C=2.5\times 10^{-11} $F, $ L=2\times 10^{-10} $.}
	\label{fig2}
\end{figure}

The model where a giant atom couples to the LHSM is depicted in Fig.~\ref{fig1},  it can be regarded as a one-dimensional waveguide. The LHSM consists of two alternating left-handed inductor-capacitor (LC) cells, each formed by series capacitors and grounded inductors \cite{PhysRevA.99.032325}. The ratio of capacitance (inductance) between neighboring cells is denoted as $ \epsilon $. The length of one LC cell is designed as $ \Delta x $. We consider two adjacent LC cells as a superlattice unit with a length of $ \Delta X=2\Delta x $. The Lagrangian of the LHSM is~\cite{PhysRevLett.111.163601,PhysRevA.69.062320}

\begin{eqnarray}
\mathcal{L}&=&\frac{1}{2}\sum_n{\left[ C( \dot{\Phi}_n-\dot{\Phi}_{n-1} ) ^2+\epsilon 	C( \dot{\Phi}_n-\dot{\Phi}_{n+1}) ^2 \right]}  \notag \\
 &-&\frac{1}{2}\sum_n{\left[ \frac{1}{\epsilon L}\Phi _{n}^{2}+\frac{1}{L}\Phi _{n-1}^{2} \right]},
\label{lagranian}
\end{eqnarray}
with  $ \textit{C} $  ($ \textit{L} $) represents the capacitance (inductance) of the LHSM. Assuming that the field takes the form of a plane wave, denoted as $ \Phi _n=e^{i\left( kn\Delta x-\omega t \right)} $, we obtain the dispersion relation of the LHSM by deriving the Euler-Lagrange equation (see details in Appendix \ref{Appendix A})
\begin{gather}
\omega_\pm =\frac{\omega _{r}}{\sqrt{\frac{\left( 1+\epsilon \right) ^2}{2}\pm \sqrt{\frac{\left( 1+\epsilon \right) ^4}{4}+\epsilon ^2\left[ 2\cos \left( k\Delta X \right) -2 \right]}}},
\label{omega}
\end{gather}
where the resonance frequency of an individual LC cell is denoted as $ \omega _r={{1}/{\sqrt{CL}}} $, with $ k $ being the wave vector. In our study, we set the values of $ C=2.5\times 10^{-11} $F, $ L=2\times 10^{-10} $H, respectively. Under these conditions, we plot the dispersion relation $ \omega_\pm(k )$ as a function of $k$, as shown in  Fig.~\ref{fig2}(a), while taking  $\epsilon =1.4$. We find that for $ k=0 $, the upper band exhibits divergence, while the lower band converges toward the infrared cutoff frequency. As $k$ increases, the frequency $\omega_+(k)$ gradually decreases to a finite value, corresponding to the left-hand characteristic inherent in this model. Simultaneously, due to the Bragg scattering, $\omega_-(k)$ increases to a finite value. The resulting band gap, $[\omega_{-}(\pm\pi),\omega_{+}(\pm\pi)] $, displays asymmetry arsing from distinct underlying mechanisms.

In Fig.~\ref{fig2}(b), we plot the relationship between the superlattice parameter $\epsilon$ and two important quantities: the band gap width $ \Delta_G $ and the width of lower band $ W_- $, i.e.,
\begin{eqnarray}
\Delta_G=\omega_+(\pm \pi)-\omega_-(\pm \pi), \quad W_-=\omega_-(\pm \pi)-\omega_-(0).
\end{eqnarray}
The LHSM is constructed from two periodic substructures with distinct refractive indices. Within the system, the band gap arises as a result of destructive interference in Bragg scattering occurring at the interface of cell $ \rm{a} $ and $ \rm{b} $. Specifically, when $ \epsilon=1 $, the band gap reaches its maximum width. In this case, all cells have the same index of refraction, rendering the LHSMis isotropic. Therefore, the band gap disappears due to lack of Bragg scattering, resulting in a band gap width of zero. The phenomenon has been previously investigated in Ref.~\cite{PhysRevApplied.11.054062}. When  $ \epsilon $ deviates from $ \epsilon=1 $, the difference in the refractive indices between neighboring cells increases. This amplifies the strength of Bragg scattering at cell boundaries. In this work, for the sake of generality, we take the superlattice parameter $ \epsilon=1.4 $.

\section{Giant atom interacting with LHSM}
\label{section:2}

As shown in Fig.~\ref{fig1}, the giant atom interacts with the LHSM at two distinct points through capacitances ~\cite{PhysRevApplied.14.064033,PhysRevLett.126.043602,PhysRevResearch.4.023198}. The giant atom takes the form of, for example, a transmon qubit consisting of two identical Josephson junctions. The Hamiltonian of the transmon qubit can be expressed in terms of the charge operator $ \hat{n} $ and the phase operator $ \hat{\varphi} $ ~\cite{PhysRevA.76.042319,Wang_2022,PhysRevX.11.011015,PhysRevA.103.023710,Calzona_2023,PhysRevA.106.042605,GU20171}
	\begin{gather}
		\hat{H}_T=4E_C \hat{n} ^2-E_J\cos \hat{\varphi},\notag \\
		\hat{n}={\hat{Q}}/{2e}, \quad \hat{\varphi}=\left( 2\pi /\varPhi _0 \right) \hat{\varPhi},
		\label{energy}
	\end{gather}
	where $ E_{J}\,\,[ E_C=e^2/(2C_{\Sigma}) ]  $ represents the Josephson (charging) energy of the transmon. The total capacitance is $ C_{\Sigma}=C_{J}^{q}+2C_q $.
	
	For transmon qubit, since $ E_J/E_C \gg  1 $, the charge zero-point fluctuations dominate over the phase zero-point fluctuations, i.e., $ \sigma \left( \hat{n} \right) \gg \sigma \left( \hat{\varphi} \right) $. Therefore, we express the transmon Hamiltonian as
	\begin{gather}
		\hat{H}_T=4E_C\hat{n}^2+\frac{1}{2}E_J\hat{\varphi}^2-E_J\left( \cos \hat{\varphi}+\frac{1}{2}\hat{\varphi}^2 \right),
		\label{H_t}
	\end{gather}	
	where we have neglected the bias charge term associated with $ n_g $ \cite{RevModPhys.93.025005,PhysRevA.76.042319}. Due to $ \sigma \left( \hat{\varphi} \right) \ll 1  $, Eq.~(\ref{H_t}) can be rewritten as
	\begin{gather}
		\hat{H}_q=4E_C\hat{n}^2+\frac{1}{2}E_J\hat{\varphi}^2-\frac{1}{4!}E_J\hat{\varphi}^4,
		\label{H_q}
	\end{gather}
	which can be viewed as a Duffing oscillator \cite{RevModPhys.93.025005,PhysRevA.76.042319,Krantz2019}. The charge operator and phase operator can be denoted by the creation and annihilation operator  $ \hat{b}^{\dagger} $ and $ \hat{b} $ with
	\begin{gather}
		\hat{\varphi}=\left( \frac{2E_C}{E_J} \right) ^{\frac{1}{4}}( \hat{b}^{\dagger}+\hat{b} ), \quad
		\hat{n}=\frac{i}{2}\left( \frac{E_J}{2E_C} \right) ^{\frac{1}{4}}( \hat{b}^{\dagger}-\hat{b} ).
	\end{gather}
	Therefore, the Hamiltonian can be expressed in a simplified form
	\begin{eqnarray}
		\hat{H}_q&=&\sqrt{8E_CE_J}\hat{b}^{\dagger}\hat{b}-\frac{E_C}{12}\left( \hat{b}^{\dagger}+\hat{b} \right) ^4  \notag \\
		&\approx& \hbar \omega _q\hat{b}^{\dagger}\hat{b}-\frac{E_C}{2}\hat{b}^{\dagger}\hat{b}^{\dagger}\hat{b}\hat{b},
	\end{eqnarray}
	with $ \omega _q=\sqrt{8E_CE_{J}^{q}}-E_C $.
	
	When the coupling strength is significantly smaller than the anharmonicity of the transmon qubit, i.e., $ g\ll \eta \simeq -E_C  $, we can neglect higher energy levels and consider the transmon qubit as a two-level system. By considering the two lowest energy levels of the emitter, we employ transformations $ \hat{b}^{\dagger}\hat{b}\to\sigma _z,\;\hat{b}\to \sigma _-,\;\hat{b}^{\dagger}\to \sigma _+ $ to describe the system Hamiltonian \cite{Krantz2019,PhysRevA.76.042319,RevModPhys.93.025005}, i.e.,
	\begin{gather}
		H_q=\frac{1}{2}\omega _q\sigma _z.
\end{gather}

As derived in Refs.~\cite{PhysRevB.92.104508,PhysRevLett.126.043602}, the Hamiltonian of the LHSM can be quantized as (see details in Appendix \ref{Appendix B})
\begin{gather}
\hat{H}_0=\sum_{k=1}^N{\hbar \omega _k\left( a_{k}^{\dagger}a_k+\frac{1}{2} \right)},
\end{gather}
where $ a_k (a_{k}^{\dagger})  $ is the annihilation (creation) operator of the photonic modes with the wave vector $ k $.

In the rotating-wave approximation, the interaction Hamiltonian between transmon qubit and the LHSM is expressed as
\begin{gather}
H_{\mathrm{int}}=\sum_k{g_k( \hat{a}_{k}^{\dagger}\hat{\sigma}_-+\hat{a}_k\hat{\sigma}_+ )},
\end{gather}
where $ \sigma _+=( \sigma _- ) ^{\dagger}=|e\rangle \langle g| $, with $ |e\rangle ( \langle g| )  $ being the excited (ground) state of the emitter. 
The coupling strength is given by \cite{PhysRevLett.126.043602}
\begin{gather}
g_k=\frac{e}{\hbar}\frac{C_{J}^{g}}{C_{\Sigma}}\sqrt{\frac{\hbar \omega _k}{C_W}},
\end{gather}
with $ C_W $ denoting the total capacitance of the LHSM waveguide.
Finally, by setting $\hbar=1$, the Hamiltonian of the system can be described by
\begin{gather}
\hat{H}=H_0+H_q+H_{\mathrm{int}}
=\frac{1}{2}\omega _q\sigma _z   \notag   \\
+\sum_k{\omega _ka_{k}^{\dagger}a_k}
+\sum_k{g_k( \hat{a}_{k}^{\dagger}\hat{\sigma}_-+\hat{a}_k\hat{\sigma}_+ )}.
\label{total Hamiltonian}
\end{gather}

\section{The dynamics of the system}
\subsection{Quantum dynamics in the dispersive band}
\label{section:3}
When the emitter resonates with the upper (lower) band, there will be a significant number of modes with non-zero group velocity coupled to the emitter. This coupling phenomenon leads to an exponential emission of photons by the emitter. However, as the emitter's frequency approaches the band edge, the Wigner-Weisskopf approximation will break down, leading to the non-Markovian dynamics \cite{RevModPhys.87.347,PhysRevA.106.043703,Andersson_2019}. We first explore the spontaneous decay of the giant atom when its frequency is significantly removed from the band edges.

In the rotating frame of atomic frequency $ \omega_{q} $, the total Hamiltonian, as given in Eq.~(\ref{total Hamiltonian}), is derived as \cite{PhysRevLett.78.2539}
\begin{eqnarray}
H=\sum_{k\in \mathrm{BZ}}{\Delta _ka_{k}^{\dagger}a_k+\sum_{k\in \mathrm{BZ}}{( g_ka_{k}^{\dagger}\sigma _-+g_{k}^{\ast}a_k\sigma _+ )}},
\end{eqnarray}
with the frequency detuning is $ \Delta _k=\omega _k-\omega _q $ (within the first Brillouin zone (BZ)). The system's state can be expanded in the single excitation subspace as
\begin{gather}
|\psi \left( t \right) \rangle =\sum_k{c_{g,k}\left( t \right) |g,1_k\rangle}+c_e\left( t \right) |e,0\rangle,
\end{gather}
here $ |g,1_k\rangle  $ corresponds to the state where the giant atom is in the ground state, and a single photon is excited at mode $ k $. We assume that the giant atom (waveguide) is initially in the excited (vacuum) state, i.e., $ |\psi \left( t=0 \right) \rangle = |e,0\rangle $. According to Schrodinger equation, we obtain the following differential equations
\begin{eqnarray}
&&\dot{c}_{g,k}\left( t \right) =-i\left[ \Delta _kc_{g,k}\left( t \right) +g_kc_{e}\left( t \right) \right], 
\label{cgk}\\
&&\dot{c}_{e}\left( t \right) =-i\sum_k{g_{k}^{\ast}c_{g,k}\left( t \right)}.
\label{cek}
\end{eqnarray}
By defining $ \tilde{c}_{g,k}\left( t \right) =c_{g,k}\left( t \right) e^{i\Delta _kt} $ and substituting its integral form into Eq.~(\ref{cek}), we obtain
\begin{gather}
\dot{c}_{e}\left( t \right) =\sum_k{g_{k}^{2}\int_0^t{c_{e}\left( t' \right) e^{i\Delta _k\left( t-t' \right)}\mathrm{d}t'}}
\label{ce_evolution}.
\end{gather}

Note that $ g_{k} $ is the coupling strength in $ k $ space \cite{PhysRevLett.126.043602}. We consider the giant atom coupling to the waveguide at two points $ x_1=0 $ and $ x_2=d_s $. The separation distance $ d_s $ corresponds to the giant atom's size. Unlike the setup with small atom where $ g_{k} $ is a constant, the coupling strength $ g_k $ for giant atoms exhibits dependence on the parameter $ d_s $, i.e.,
\begin{eqnarray}
g_k=g\left( 1+e^{ikd_s} \right).
\end{eqnarray}

 The summation over $ k $ can be replaced with an integral, i.e., $ \sum_k{\rightarrow \frac{N}{2\pi}\int_{-\pi}^{\pi}{\mathrm{d}k}} $. We can rewrite Eq.~(\ref{ce_evolution}) as
	\begin{gather}
		\dot{c}_e\left( t \right) =-\frac{N}{2\pi}\int_{-\pi}^{\pi}{g_{k}^{2}\mathrm{d}k\int_0^t{c_e\left( t' \right) e^{i\Delta _k\left( t-t' \right)}}\mathrm{d}t'}.
	\end{gather}
	
	We consider that the emitter is resonant with the upper (lower) band at $ k_r\,\,\left( k_r>0 \right) $, i.e.,  $  \omega_{q}=\omega_{k_{r}} $. As depicted in Fig.~\ref{fig2}, since the resonant frequency is significantly separated from the band edges, the dispersion relation around $ k $ can be approximated as a linear relation, i.e., $ \omega _k \simeq v_g(k)k $, with $ v_g(k) $ being the group velocity at $ k $. By calculating $ v_{g}^{\pm}(k)={\mathrm{d}\omega _{\pm}\left( k \right)}/{\mathrm{d}k} $, we obtain the group velocity $ v_g $
	
	\begin{widetext}
		\begin{equation}
			v_{g}^{\pm}(k)=\frac{-\epsilon ^2\sin k}{2\sqrt{\frac{\left( \epsilon +1 \right) ^4}{4}+\epsilon ^2\cdot \left( 2\cos k-2 \right)}\left[ \frac{\left( \epsilon +1 \right) ^2}{2}\mp \sqrt{\frac{\left( \epsilon +1 \right) ^4}{4}+\epsilon ^2\cdot \left( 2\cos k-2 \right)} \right] ^{\frac{3}{2}}},
		\end{equation}
	\end{widetext}
	where the group velocity $ v^{+(-)}_{g}(k) $ of the upper (lower) band is of the left-handed characteristic. In the Born-Markovian regime, the coupling strength is much smaller than the bandwidth around $ k $, allowing us to extend the integral $ \pm \pi  $ bound to be infinite. Moreover, in the emission spectrum, the atomic spontaneous radiation is centered on the transition frequency $ \omega _q $, we can employ the integral $\int_{-\infty}^{\infty}{d\omega _ke^{i\left( \omega _k-\omega _q \right) \left( t-t' \right)}=2\pi \delta \left( t-t' \right)} $ and replace $ [1+\cos \left( kd_s \right)]/v_g(k)  $ by $ [1+\cos \left( k_rd_s \right)]/v_{g}( k_r)  $ \cite{PhysRevA.92.023825,Ghafoor_2014,GLAETZLE2010758}. Consequently, the evolution $ \dot{c}_e(t) $ is derived as 
	\begin{gather}
		\dot{c}_e\left( t \right) =-\frac{N}{2\pi}\int_{-\infty}^{\infty}{\frac{g_{k}^{2}}{v_g(k)}e^{i\Delta _k\left( t-t' \right)}d\omega _k}\int_0^t{c_e\left( t' \right) dt'}
		\\
		=-\frac{N}{2\pi v_{k_r}}\left( g_{kr}^{2}+g_{-kr}^{2} \right)  \int_0^t{2\pi \delta \left( t-t' \right) c_e\left( t' \right) dt'}.
	\end{gather}
	where $v_{k_r}$ is the group velocity at $ k_r$ ($k_r>0$). Consequently, the probability amplitude $ c_e(t) $ is derived as
	\begin{gather}
		\dot{c}_e\left( t \right) =-\frac{2Ng^2}{v_{k_r}}[1+\cos \left( k_rd_s \right)]c_e\left( t \right).
	\end{gather}

\begin{figure*}[ht!]
	\centering\includegraphics[width=15cm]{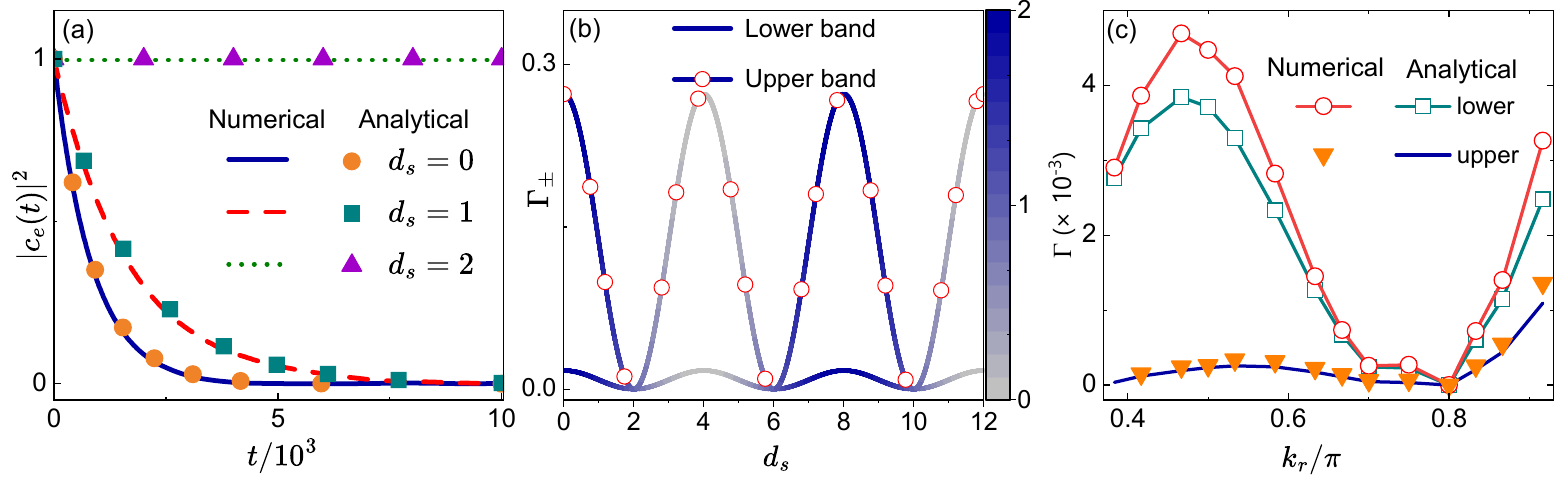}
	\caption{(a) The spontaneous decay rate of the giant atom changes with giant atom's size $ d_{s} $. We fix $ \omega_q=\omega_+ (k=\pi/2) $. The curve and dots correspond to the numerical and analytical results, respectively. (b) Dynamical evolution obtained via numerical simulation for various $ d_{s} $. (c) The spontaneous decay rate of a giant atom resonating with mode $ k_{r} $ of the lower (upper) band. The coupling strength is set as $ g=0.0001 $. Other parameters remain consistent with those in Fig.~\ref{fig2}.}
	\label{fig3}
\end{figure*}

We solve the equation for $ c_e\left( t \right) $ under the Weisskopf-Wigner approximation and obtain \cite{scully_zubairy_1997}
\begin{gather}
c_e\left( t \right) = e^{-\frac{\Gamma}{2}t},\quad \Gamma =-\frac{4Ng^2}{v_{k_r}}[ 1+\cos \left( k_rd_s \right)],
\label{gamma}
\end{gather}
where $ \Gamma $ is the spontaneous decay rate of the giant atom. Note that $ \Gamma $ depends on the size of the giant atom $ d_{s} $. These approximations are valid in our work, and can rigorously display the underlying dynamics. As depicted in Fig.~\ref{fig3}(a), the derived analytical results  with Markovian approximation are in excellent agreement with the numerical calculations.

By setting $ \omega_q=\omega(k_{r}=\pi/2) $, we depict the spontaneous decay rate as a function of the giant atom's size in Fig.~\ref{fig3}(b). The simulation methods can be found in Appendix C. The color-coding of the curves corresponds to the varying coupling strengths. It can be verified from Eq.~(\ref{gamma}) that the spontaneous decay rate exhibits periodic behavior in response to changes in the emitter's size. Given that $ \Gamma=0 $, when $ d_s=2M $ with M being an odd integer, the emitter is trapped in its excited state without decaying. We demonstrate the decay rates, calculated through dynamic evolution for various values of $ k_r $, in Fig.~\ref{fig3}(c). Regarding the calculation of the spontaneous decay rate, we initially assumed that the dispersion relation near $ k_r $ approximates linearity, i.e., $ \omega_k \simeq v_g k $. However, the dispersion is nonlinear in fact [see Fig.~\ref{fig2}(a)], leading to discrepancies between our numerical and analytical fittings. Furthermore, the presence of asymmetric energy bands gives rise to distinct spontaneous decay dynamics when the atom couples to its respective continuum. At the coupling position $k_r$, the spontaneous decay rate $ \Gamma $ within the lower energy band greatly exceeds that within the upper band due to the substantial disparity in group velocities.

\subsection{Quantum dynamics in the asymmetric band gap}
\label{section:4}
In this work, we explore the behavior of bound state of a single giant atom by considering $ \omega_{q} $ inside the asymmetric band gap \cite{RN130}. There is no continuum mode resonant with the giant emitter. As a result, spontaneous emission is suppressed, leading to the confinement of energy in the form of a bound state \cite{PhysRevA.93.033833,PhysRevA.106.043703,PhysRevA.102.033706}.

To derive the evolution analytically, we utilize the Laplace transform
\begin{gather}
\tilde{c}_{g,k\left( e \right)}\left( s \right) =\int_0^{\infty}{c_{g,k\left( e \right)}\left( t \right) e^{-st}\mathrm{d}t},
\end{gather}
Eqs.~(\ref{cgk},\ref{cek}) are respectively derived as ~\cite{PhysRevA.106.043703}
\begin{eqnarray}
&&s\tilde{c}_e\left( s \right) -s\tilde{c}_e\left( 0 \right) =-i\sum_k{g_k\tilde{c}_{g,k}\left( s \right)},
\label{ces}\\
&&s\tilde{c}_{g,k}\left( s \right) -s\tilde{c}_{g,k}\left( 0 \right) =-i\Delta _k\tilde{c}_{g,k}\left( s \right) -ig_k\tilde{c}_e\left( s \right) .
\label{cks}
\end{eqnarray}

\begin{figure*}[ht!]
	\centering\includegraphics[width=15cm]{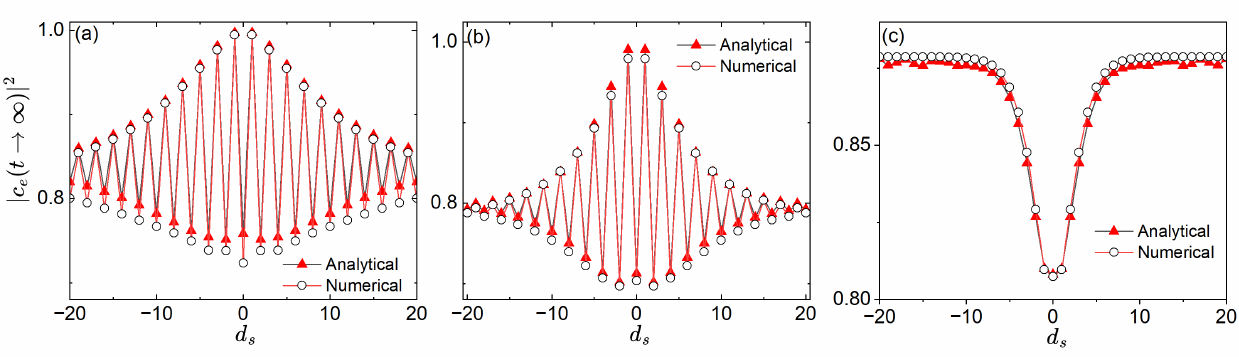}
	\caption{Bound states of giant atoms. The population of trapped atomic states $ \left| c_e\left( t\rightarrow \infty \right) \right|^2 $ varies as a function of the giant atom's size $ d_{s} $ for different conditions: (a) $ \omega_q=\omega_1 $,  (b) $ \omega_q=\omega_2 $, (c) $ \omega_q=\omega_3 $. The parameters are the same as those in Fig.~\ref{fig2} and Fig.~\ref{fig3}. The solid line marked by the circle and triangle dots represents the numerical and analytical solution,  respectively. }
	\label{fig4}
\end{figure*}

Under the initial condition $ c_e\left( 0 \right) =1 $ and $ c_{g,k}\left( 0 \right) =0 $, Eq.~(\ref{cks}) can be simplified as
\begin{gather}
\tilde{c}_{g,k}\left( s \right) =\frac{-ig_k\tilde{c}_e\left( s \right)}{\left( s+i\Delta _k \right)}.
\label{ssce}
\end{gather}
By substituting Eq.~(\ref{ssce}) into Eq.~(\ref{ces}), we obtain ~\cite{PhysRevA.96.043811,R100}
\begin{eqnarray}	
\,\,\tilde{c}_e\left( s \right) =\frac{1}{s+\sum{_e\left( s \right)}}\,\, ,
\label{ces1}\\
\sum_e{\left( s \right)}=\sum_k{\frac{|g_k|^2}{s+i\Delta _k}},
\label{self energy}
\end{eqnarray}
where $ \sum_e{\left( s \right)} $ is the so-called self-energy.
Then we can take the inverse Laplace transform of Eq.~(\ref{ces1}) in the complex space to get the time-dependent evolution $ c_e\left( t \right)  $ and obtain
\begin{gather}
c_e\left( t \right) =\frac{1}{2\pi i}\underset{E\rightarrow \infty}{\lim}\int_{\gamma -iE}^{\gamma +iE}{\tilde{c}_e\left( s \right) e^{st}\mathrm{d}s},
\end{gather}
where $ \gamma $ ($ \gamma>0 $) is a real number that makes the path integral of $ \tilde{c}_{g,k\left( e \right)}\left( s \right)  $ in the domain of convergence.
As depicted in Fig.~\ref{fig4}(a), we assume the emitter’s frequency to be $ \omega_q=\omega_3 $ [refer to Fig.~\ref{fig2}], and only the modes with $ k=0 $ contribute significantly to the system's dynamics. When the frequency resides within the asymmetric band gap, denoted as $ \omega_q=\omega_1 $ ($ \omega_2 $), we confine our analysis to modes around $ k=\pi $. Consequently, around $ k=0 $ or $ k=\pi $, the dispersion relation can be effectively approximated by a quadratic function, i.e.,
\begin{equation}
\begin{cases}
E_+\left( k \right) =E_{+\min}+\alpha_{+} \left( k\pm \pi \right) ^2,\\
E_-\left( k \right) =E_{-\max}-\alpha_{-} \left( k\pm \pi \right) ^2,\\
E_-\left( k \right) =E_{-\min}+\alpha_{0} \left( k-0 \right) ^2,\\
\end{cases}\quad \begin{array}{c}
\omega _q=\omega_1,\\
\omega _q=\omega_2,\\
\omega _q=\omega_3,\\
\end{array}
\label{Ek_q}
\end{equation}

At the band edges, we denote the curvature $ \alpha_{\pm} $ and $ \alpha_{0} $ as the second-order derivatives, which is expressed as
\begin{equation}
\alpha_{\pm}=\frac{\mathrm{d}^2E_{\pm}\left( k \right)}{\mathrm{d}k^2}\Big|_{k=\pm k_{0}}.
\label{alpha}
\end{equation}
In this case, by setting $ \delta k=k-k_0 $, the interaction strength is written as
\begin{eqnarray}
g_k=g\left[ 1+e^{id_s\left( k_0+\delta k \right)} \right].
\label{gk}
\end{eqnarray}
By replacing $ \sum_k $ as the integral form $ \frac{N}{2\pi}\int{\mathrm{d}k} $, we rewrite Eq.~(\ref{self energy}) as
\begin{eqnarray}
\sum{_e\left( s \right)}\simeq \frac{N}{2\pi}\int_{-\pi}^{\pi}{\frac{\left| g_k \right|^2}{s+i\Delta _k}\mathrm{d}k}.
\label{self-energy}
\end{eqnarray}
Finally by inserting Eqs.~(\ref{Ek_q},\ref{alpha},\ref{gk}) into Eq.~(\ref{self-energy}), we obtain
\begin{eqnarray}
\sum_e{\left( s \right)}&=&\frac{Ng^2}{\pi}\Bigg\{ \int_{-\pi}^0{\frac{1+\cos \left[ d_s\left( \delta k+k_0 \right) \right]}{s+i\left[ \Delta _0+\alpha_{\pm(0)} \left( k+k_0 \right) ^2 \right]}}\mathrm{d}\left( \delta k \right)  \notag \\ &+&\int_0^{\pi}{\frac{1+\cos \left[ d_s\left( \delta k-k_0 \right) \right]}{s+i\left[ \Delta _0+\alpha_{\pm(0)} \left( k-k_0 \right) ^2 \right]}}\mathrm{d}\left( \delta k \right)
\Bigg\}  .
\end{eqnarray}

Due to the emitter's frequency is close to the edge of the upper (lower) band, we limit our consideration to modes around $ k_0=0  $ $ (k_0=\pi) $ when calculating the self-energy. As a result, the self-energy is derived as
\begin{eqnarray}
\sum_e{\left( s \right)}=\frac{-iNg^2}{\sqrt{\alpha \left( \Delta _0-x \right)}}\left[ 1+\cos \left( d_sk_0 \right) e^{-d_s\sqrt{\frac{\Delta _0-x}{\alpha}}} \right] .
\end{eqnarray}
We can use the residue theorem to obtain the steady-state probability
\begin{gather}
|c_e(t=\infty)|^2=|\text{Res}(s_{0})|^2, \\	\text{Res}(s_{0})=\frac{1}{1+\partial_s \Sigma_{e}(s)}\Big|_{s=s_{0}},
\label{res}
\end{gather}
where  Res($ s_{0} $)  is the steady population for giant atom, and $ s_{0} $ is the purely imaginary pole of the transcendental equation, which can be obtained by
\begin{gather}
s+\sum_e{\left( s_0 \right) =0}.
\end{gather}

Given that giant atom is coupled to the LHSM waveguide at two distinct points, static bound states are formed at each of these coupling locations. As the separation between these coupling points diminishes, the two bound states interfere, giving rise to a periodic interference pattern in the dynamical evolution of the giant atom. As depicted in Fig.~\ref{fig4}(a,b), we observe that the dynamics evolution of the emitter's population, $ \left| c_e\left( t \right) \right|^2 $ varies with $ d_{s} $. Due to the asymmetric nature of the band gap, the curvatures $ \alpha_{\pm} $, which correspond to different mode densities, exhibit dissimilarity. Therefore, the interference patterns at the upper (lower) band edges exhibit disparities.  When $ d_{s} $ is odd, it leads to a dominant destructive interference, causing the coupling strength to nearly vanish. Consequently, the majority of the energy remains confined within the emitter, with minimal escape into the waveguide.

Conversely, for even values of $ d_{s} $, constructive interference prevails, resulting in a significantly reduced trapped atomic population, as depicted in Fig.~\ref{fig4}(a,b). In cases where $ d_{s} $ is comparable to or excess the size of the bound state, the interference effect diminishes, and the steady-state atomic population asymptotically reaches its stable value.

\begin{figure}[ht!]
	\centering\includegraphics[width=8cm]{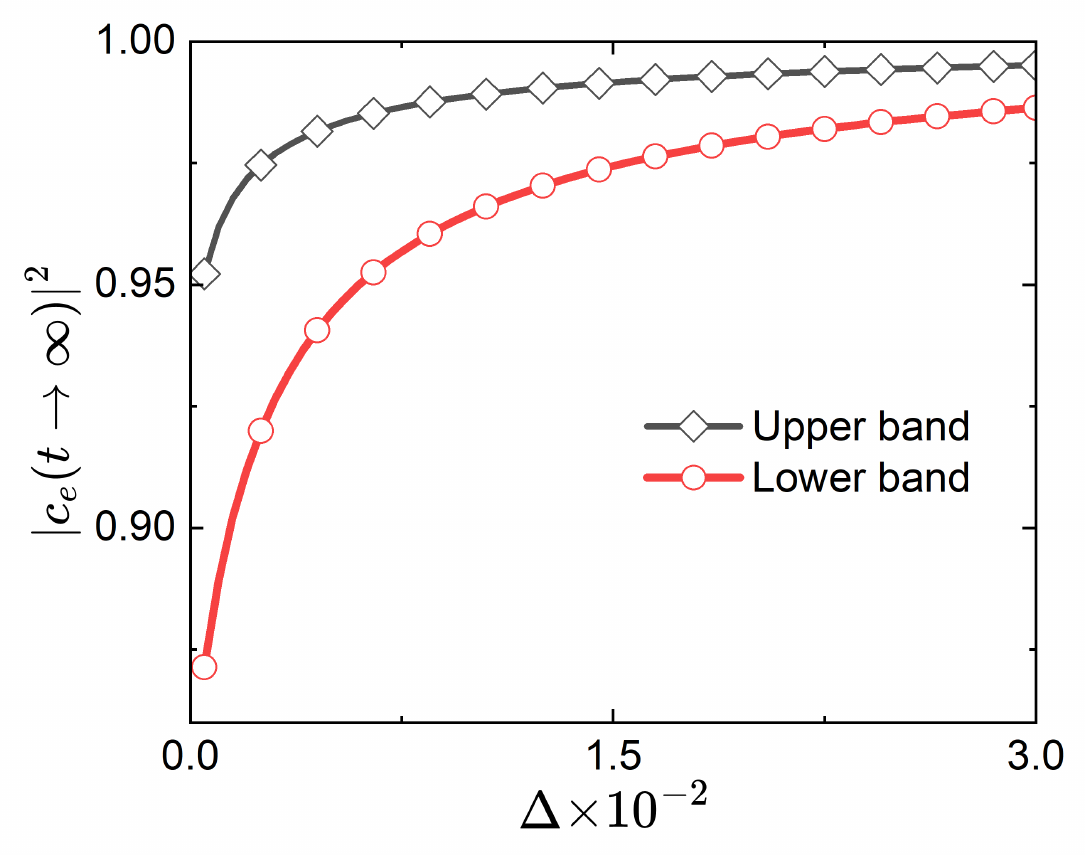}
	\caption{The trapped atomic population changes with detuning $ \Delta $ to the upper (lower) band edges, respectively. The parameters are the same as those in Fig.~\ref{fig4}.}
	\label{fig5}
\end{figure}

In Fig.~\ref{fig5}, we depict the steady-state population as a function of detuning $ \Delta $ for the upper (lower) band. It is note that due to the distinct mode densities in these two bands, the amplitude of the steady state in the upper band consistently exceeds that of the lower band. Moreover, as $ \omega_q $ is tuned towards the lower-bound of the lower band [see $ \omega_3 $ in Fig.~\ref{fig2}(a)], the oscillating interference effect no longer exists, since only the modes around $ k=0 $ are excited (satisfying the condition $ kd_{s}=0 $). In cases where the two fields do not significantly overlap for large values of $ d_{s} $, the steady-state population converges to a constant value.

\section{Two emitters}
\label{section:5}
\begin{figure}[ht!]
	\centering\includegraphics[width=8cm]{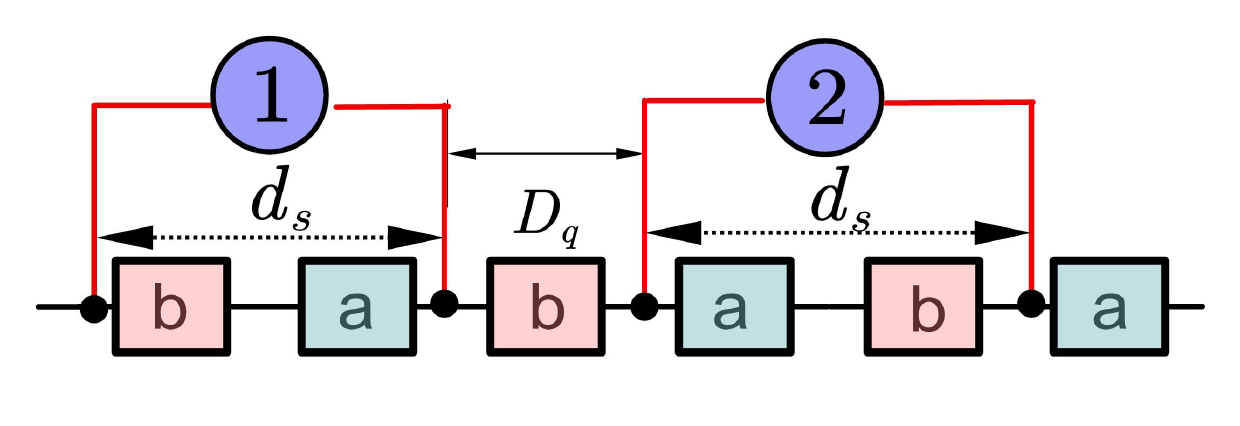}
	\caption{ Two giant emitters with sizes $ d_{s} $, coupling to the LHSM. The separation between two emitters is denoted as $ D_{q} $.}
	\label{fig7}
\end{figure}
\begin{figure*}[ht!]
	\centering\includegraphics[width=15cm]{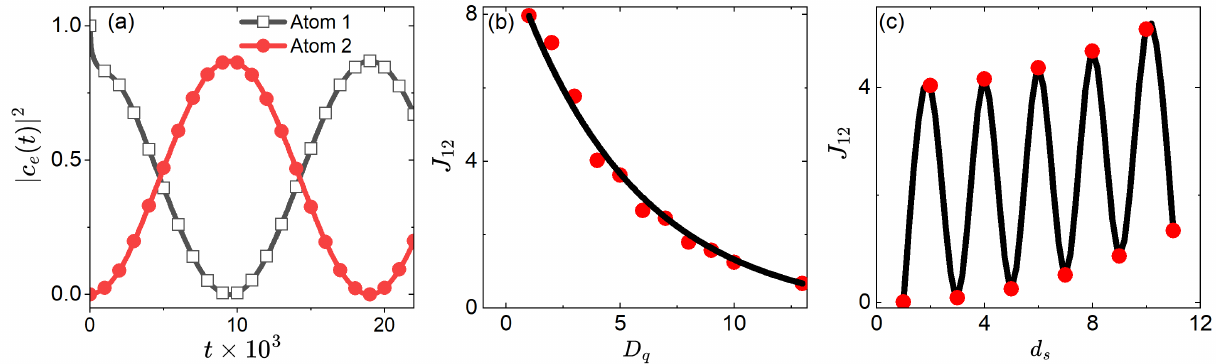}
	\caption{ (a) Rabi oscillations between two giant emitters. The parameters are set as $ d_{s}=3 $, $ D_{q}=6 $, and $ g=0.0001 $. (b) The Rabi frequency of two giant two emitters varies with  $ D_{q} $. (c) Rabi oscillations as a function of $ d_s $. Other parameters remain consistent with those in Fig.~\ref{fig4}.}
	\label{fig6}
\end{figure*}
As shown in Fig.~\ref{fig7}, we now consider two identical giant atoms interacting with the LHSM separated by a distance $ D_{q} $, with frequencies $ \omega_q $ inside the bandgap. When the separation distance $ D_q $ between atoms is relatively small, the bound states of two atoms will overlap, leading to a strong interaction between them \cite{RN231,PhysRevLett.130.053601,PhysRevLett.130.053601}. As $ D_q $ increases, the overlap area of the two fields diminishes, and the dipole-dipole interaction becomes weak. Similar to the case when a single emitter couples to the waveguide, in the rotating frame, the interaction Hamiltonian is written as
~\cite{PhysRevA.107.013710,PhysRevA.87.033831,PhysRevResearch.2.043184,PhysRevA.85.053827}
\begin{gather}
H_I=\sum_{i=1,2}{\sum_k{g_{ki}a_{k}^{\dagger}e^{i\Delta _kt}\sigma _{i}^{-}}}+H.c.
\end{gather}

At the initial state, one atom is excited and the other is in the ground state, with the basis $ |e,g,1_k \rangle  $ and $ |g,e,1_k\rangle  $. Employing the framework of effective Hamiltonian theory~\cite{effective}, the effective Hamiltonian can be expressed in the form 
\begin{eqnarray}
H_{\mathrm{eff}}\left( t \right) &=&\sum_{m,n}{\frac{1}{\bar{\omega}_{mn}}}\left[ \hat{A}_{m}^{\dagger},\hat{A}_n \right] e^{i\left( \omega _m-\omega _n \right) t},
\label{effective Hamiltonian}\\
\frac{1}{\bar{\omega}_{mn}}&=&\frac{1}{2}\left( \frac{1}{\omega _m}+\frac{1}{\omega _n} \right),
\end{eqnarray}
where $ \bar{\omega}_{mn} $ is the average of $ \omega_m $ and $ \omega_n $, with $\Delta_k=\omega_k-\omega_q  $. We make the identification $ A_{1}^{\dagger}=g_{k1}a_{k}^{\dagger}\sigma _{1}^{-} $ , $ A_{2}^{\dagger}=g_{k2}a_{k}^{\dagger}\sigma _{2}^{-} $. Substituting these conditions into Eq.~(\ref{effective Hamiltonian}), we obtain the system's effective Hamiltonian as

\begin{eqnarray}
	H_{\mathrm{eff}}= \sum_{i=1,2}\sum_k{\frac{g_{ki}g_{ki}^{\ast}}{\Delta _k}\left( \sigma _{i}^{-}a_{k}^{\dagger}\sigma _{i}^{+}a_{k}-\sigma _{i}^{+}a_{k}\sigma _{i}^{-}a_{k}^{\dagger} \right)} \notag \\ +\sum_k{\frac{g_{k1}g_{k2}^{\ast}}{\Delta _k}\left( \sigma _{1}^{-}a_{k}^{\dagger}\sigma _{2}^{+}a_k-\sigma _{2}^{+}a_k\sigma _{1}^{-}a_{k}^{\dagger} \right)}+ \mathrm{H.c.}
\end{eqnarray}
 The first two terms correspond to the atomic frequency shift, while the second and third pair of terms account for the exchange interaction between the two atoms. Since the two emitters are alternately excited, the waveguide can be approximated to be in the vacuum state. Therefore, under the approximation
\begin{gather}
\langle a_{k}^{\dagger}a_k \rangle \simeq 0,\quad \langle a_ka_{k}^{\dagger} \rangle \simeq 1.
\end{gather}
The dipole-dipole interaction Hamiltonian can be simplified as
\begin{gather}
H_{\mathrm{eff},\mathrm{d}}=-\sum_k{\frac{g_{k1}g_{k2}^{\ast}}{\Delta _k}\sigma _{2}^{+}a_k\sigma _{1}^{-}a_{k}^{\dagger}}+\mathrm{H}.\mathrm{c}.
\end{gather}

We can obtain the interaction strength
\begin{gather}
J_{12}=\sum_k{\frac{ g_{k1}g_{k2}^{\ast} }{\Delta _k}},
\label{Jab}
\end{gather}
where the coupling strengths of two giant atoms are respectively given as
\begin{gather}
g_{k1}=g\left( 1+e^{ikd_s} \right),\quad g_{k2}=g_{k1}e^{ikD_q}.
\label{gk of two}
\end{gather}
Substituting Eq.~(\ref{gk of two}) into Eq.~(\ref{Jab}) and replacing the sum with integral form, we obtain
\begin{gather}
J_{12}=\frac{N}{2\pi}\int_{-\pi}^{\pi}{\frac{2g^2[ 1+\cos \left( kd_s \right) ] e^{ikD_q}}{\Delta _k}}\mathrm{d}k,
\label{J12-1}
\end{gather}
which can be expressed as
\begin{eqnarray}
J_{12}&=&\frac{Ng^2}{\pi}\Bigg\{ \int_{-\pi}^0{\frac{\cos \left( kD_q \right) +\cos \left( kD_q \right) \cos \left( kd_s \right)}{\Delta _0+\alpha \left( k+k_r \right) ^2}}\mathrm{d}k   \notag \\
&+&\int_0^{\pi}{\frac{\cos \left( kD_q \right) +\cos \left( kD_q \right) \cos \left( kd_s \right)}{\Delta _0+\alpha \left( k-k_r \right) ^2}}\mathrm{d}k
\Bigg\} ,
\end{eqnarray}
where the dispersion relation is approximated as a quadratic form. Finally, we derive the dipole-dipole interaction strength as
\begin{gather}
J_{12}=\frac{Ng^2}{\alpha \beta}\cos \left( D_q\pi \right) e^{-D_q\beta}\left[ 1+\frac{\cos \left( d_s\pi \right)}{2}\left( e^{-d_s\beta}+e^{d_s\beta} \right) \right],  \notag \\
\beta =\sqrt{\frac{\Delta _0}{\alpha}}.
\label{dipole-dipole}
\end{gather}

In Fig.~\ref{fig6}(a), we depict the dynamics of the two emitters through numerical simulations and observe that the two atoms can coherently exchange excitation without decaying. Subsequently, Figure~\ref{fig6}(b) shows a numerical depiction of the variation of $ J_{12} $ with the separation distance $ D_{q} $, which is approximately described by an exponential form in Eq.~(\ref{dipole-dipole}). Finally, Figure~\ref{fig6}(c) demonstrates the size-dependent characteristic of the dipole-dipole interaction, which arises from the periodic modulation of the bound state by the giant atom's size $ d_s $.
\section{Conclusion}
 In this paper, we explore the quantum dynamics by considering giant atoms interacting with LHSM. The emergence of an asymmetric band gap by the left-handed dispersion band and the Bragg scattering band leads to several unconventional phenomena in quantum optics. We consider the giant atom in resonance with either the upper or lower band. Through the analysis of the interference phenomena induced by these giant atoms, we find that the spontaneous decay rate changes periodically with the giant atom's size. The asymmetric band structure results in distinct quantum dynamics for giant atoms resonating with different bands. As a consequence, spontaneous emission can be modulated by adjusting either the size or the resonant frequency of the giant atom, allowing us to enhance or suppress the process as needed.

Most remarkably, when confining the emitter's frequency within the asymmetric band gap, we find that the dynamics dramatically depends on the band edge's properties. By calculating the steady population, we observe a periodic modulation in the dynamical evolution, a consequence of the interference effects caused by variations in the giant atom's size. Moreover, the asymmetric band edges lead to different interference amplitudes for the upper (lower) band edges. Similarly, the dipole-dipole interaction between two giant atoms depends on their respective sizes and distance that separates them. This mechanism reveals that our work provides a method to engineer the interaction between giant atoms and metamaterial environment in future studies.

\section{Acknowledgments}
The quantum dynamical simulations are based on open source code 
QuTiP ~\cite{Johansson12qutip,Johansson13qutip}. 
X.W.~is supported by 
the National Natural Science
Foundation of China (NSFC) ( No.~12174303 and Grant No.~11804270), and the Fundamental 
Research Funds for the Central Universities (No. xzy012023053). W.X.L. is supported by the Natural Science Foundation of Henan Province (No. 222300420233).
\appendix
\section{Deriving the dispersion relation of the Left-handed superlattice metamaterial}
\label{Appendix A}
In this Appendix, we derive the dispersion relation of the LHSM. The Lagrangian form is given in main text Eq. (\ref{lagranian}). The structure of the LHSM is shown in Fig.~\ref{fig3}. According to Euler-Lagrange equation $ \frac{d}{dt}\frac{\partial \mathcal{L}}{\partial \dot{\phi}}-\frac{\partial \mathcal{L}}{\partial \phi}=0 $, we obtain the following equations  for motions
\begin{eqnarray}
\epsilon C\left[ \ddot{\Phi}_{n+1}-\ddot{\Phi}_n \right] -C\left[ \ddot{\Phi}_n-\ddot{\Phi}_{n-1} \right] =\frac{1}{\epsilon L}\Phi _n, 
\label{motion1}\\
C\left[ \ddot{\Phi}_{n+2}-\ddot{\Phi}_{n+1} \right] -\epsilon C\left[ \ddot{\Phi}_{n+1}-\ddot{\Phi}_n \right] =\frac{1}{L}\Phi _{n+1}.
\label{motion2}
\end{eqnarray}
By adopting Helmholtz equation $ \ddot{\Phi}=-\omega ^2\Phi $, we rewrite Eq.~(\ref{motion1}) and Eq.~(\ref{motion2}) as
\begin{eqnarray}
\omega ^2C\left[ \left( \Phi _n-\Phi _{n-1} \right) +\epsilon \left( \Phi _n-\Phi _{n+1} \right) \right] =\frac{1}{\epsilon L}\Phi _n,
\label{lagrangian_1}\\
\omega ^2C\left[ \epsilon \left( \Phi _{n+1}-\Phi _n \right) +\left( \Phi _{n+1}-\Phi _{n+2} \right) \right] =\frac{1}{L}\Phi _{n+1},
\label{lagrangian_2}
\end{eqnarray}
which result in
\begin{eqnarray}
\Phi _{n+1}=\frac{\epsilon \Phi _n+\Phi _{n+2}}{\left( \epsilon +1-\frac{1}{\omega ^2LC} \right)},\quad \Phi _{n-1}=\frac{\epsilon \Phi _{n-2}+\Phi _n}{\left( \epsilon +1-\frac{1}{\omega ^2LC} \right)}.
\label{phi}
\end{eqnarray}
After substituting Eq.~(\ref{phi}) into Eq.~(\ref{lagrangian_2}), we obtain
\begin{eqnarray}
&C&\left( 1+\epsilon -\frac{1}{\omega ^2\epsilon LC} \right) \Phi _n-C^2\frac{\epsilon \Phi _{n-2}+\Phi _n}{\left[ \left( \epsilon +1 \right) C-\frac{1}{\omega ^2L} \right]} \notag \\
&&-\epsilon C^2\frac{\epsilon \Phi _n+\Phi _{n+2}}{\left[ \left( \epsilon +1 \right) C-\frac{1}{\omega ^2L} \right]}=0.
\end{eqnarray}

By adopting the plane-wave form $ \Phi _n=e^{i\left( \textbf{k}n\Delta x-\omega t \right)} $ ($ \Delta X=2\Delta x $), the dispersion relation can be derived as 
\begin{gather}
\omega =\frac{\omega _{sl}}{\sqrt{\frac{\left( 1+\epsilon \right) ^2}{2}\pm \sqrt{\frac{\left( 1+\epsilon \right) ^4}{4}+\epsilon ^2\left[ 2\cos \left( k\Delta X \right) -2 \right]}}}.
\end{gather}

\section{Deriving the Hamiltonian of the waveguide}
\label{Appendix B}
We now calculate the Hamiltonian of the superlattice metamaterial. The Lagrangian of the LHSM  can be written as~\cite{PhysRevB.92.104508,PhysRevLett.126.043602}
\begin{gather}
\mathcal{L} =\frac{1}{2}\vec{\dot{\Phi}}^T\hat{C}\vec{\dot{\Phi}}-\frac{1}{2}\vec{\Phi}^T\hat{L}^{-1}\vec{\Phi},
\end{gather}
where the flux vector $ \vec{\Phi} $ is
\begin{gather}
\vec{\Phi}^T=\left( \Phi _0,\Phi _1,\cdots \Phi _N \right).
\end{gather}

According to Eq.~(\ref{lagranian}) in main text, the capacitance and inductance matrices are
\begin{eqnarray}
\hat{C}=C\left( \begin{matrix}{}
1&		-1&		0&		0&		\cdots\\
-1&		\left( \varepsilon +2 \right)&		-\left( \varepsilon +1 \right)&		0&		\cdots\\
0&		-\left( \varepsilon +1 \right)&		\left( 2\varepsilon +2 \right)&		-\left( \varepsilon +1 \right)&		\cdots\\
0&		0&		-\left( \varepsilon +1 \right)&		\left( 2\varepsilon +2 \right)&		\cdots\\
\vdots&		\ddots&		\ddots&		\ddots&		\ddots\\
\end{matrix} \right)  
\end{eqnarray}
and
\begin{eqnarray}
\hat{L}^{-1}=\frac{1}{L}\left( \begin{matrix}{}
1&		0&		0&		0&		\cdots\\
0&		\frac{1+\varepsilon}{\varepsilon}&		0&		0&		\cdots\\
0&		0&		\frac{1+\varepsilon}{\varepsilon}&		0&		\cdots\\
0&		0&		0&		\frac{1+\varepsilon}{\varepsilon}&		\cdots\\
\vdots&		0&		\ddots&		\ddots&		\ddots\\
\end{matrix} \right)  .
\end{eqnarray}
Based on Euler-Lagrange equation, the Hamiltonian of LHSM is given as
\begin{gather}
H_0=\vec{Q}^T\dot{\vec{\Phi}}-\mathcal{L} =\frac{1}{2}\vec{Q}^T\hat{C}^{-1}\vec{Q}+\frac{1}{2}\vec{\Phi}^T\hat{L}^{-1}\vec{\Phi},
\end{gather}
where the charge vector is defined as $ \vec{Q}=\hat{C}\dot{\vec{\Phi}} $. As derived in Refs.~\cite{PhysRevB.92.104508,PhysRevLett.126.043602}, the Hamiltonian of the LHSM is quantized as
\begin{gather}
\hat{H}_0=\sum_{k=1}^N{\hbar \omega _k\left( a_{k}^{\dagger}a_k+\frac{1}{2}, \right)}
\end{gather}
where $ a_k $ ($ a_{k}^{\dagger}  $) is the annihilation (creation) operator of the photonic mode with wave vector $ \textit{k} $.
Note that the eigenfrequency $ \omega _k $ and the eigenvectors $ \vec{\psi}_{k}^{s}=\hat{C}^{\frac{1}{2}}\vec{\Phi} $ satisfy the equation $ \hat{C}^{-\frac{1}{2}}\hat{L}^{-1}\hat{C}^{-\frac{1}{2}}\vec{\psi}_k=\omega _{k}^{2}\vec{\psi}_k $.
\section{The numerical simulation method}
 We simulate the dynamics of the system by considering the giant emitter coupling to a N-mode LHSM metamaterial. The numerical calculations proceed through the following steps:
	
	(a) In our simulation, we discretize a total of N=5000 modes within the first BZ $k\in \left( -\pi ,\pi \right] $, which is equivalent to considering a finite waveguide of length $L=5000\lambda $ in real space. The substantial length $L$ ensures that the propagating wavepacket does not reach the boundary throughout the simulaiton.
	
	(b) In the single-excitation subspace, i.e., $ |\psi \left( t \right) \rangle =\sum_k{c_{g,k}\left( t \right) |g,1_k\rangle}+c_e\left( t \right) |e,0\rangle $, the Hamiltonian can be represented as a matrix with dimensions 2N+Q, with Q as the number of atoms. To illustrate, we take two giant atoms (Q=2) for example, and the Hamiltonian matrix is given by
	\begin{gather}
		H_{int}=\left( \begin{matrix}
			\omega _{k1}^{+}&		0&		\cdots&		0&		0&		0&		\cdots&		0&		g_{k1}&		g_{k1}\\
			0&		\omega _{k2}^{+}&		\ddots&		\vdots&		0&		0&		\cdots&		0&		g_{k2}&		g_{k2}\\
			\vdots&		\ddots&		\ddots&		0&		\vdots&		\vdots&		\ddots&		\vdots&		\vdots&		\vdots\\
			0&		\cdots&		0&		\omega _{kN}^{+}&		0&		0&		\cdots&		0&		g_{kN}&		g_{kN}\\
			0&		0&		\cdots&		0&		\omega _{k1}^{-}&		0&		\cdots&		0&		g_{k1}&		g_{k1}\\
			0&		0&		\cdots&		0&		0&		\omega _{k2}^{-}&		\ddots&		\vdots&		g_{k2}&		g_{k2}\\
			\vdots&		\vdots&		\ddots&		\vdots&		\vdots&		\ddots&		\ddots&		0&		\vdots&		\vdots\\
			0&		0&		\cdots&		0&		0&		\cdots&		0&		\omega _{kN}^{-}&		g_{kN}&		g_{kN}\\
			g_{k1}^{\ast}&		g_{k2}^{\ast}&		\cdots&		g_{kN}^{\ast}&		g_{k1}^{\ast}&		g_{k2}^{\ast}&		\cdots&		g_{kN}^{\ast}&		\omega _{qa}&		0\\
			g_{k1}^{\ast}&		g_{k2}^{\ast}&		\cdots&		g_{kN}^{\ast}&		g_{k1}^{\ast}&		g_{k2}^{\ast}&		\cdots&		g_{kN}^{\ast}&		0&		\omega _{qb}\\
		\end{matrix} \right) ,
	\end{gather}
	with $ \omega_{k_i}^{\pm} $ denoting the frequency of the upper (lower) energy band and  $ g_{k_i} $ is the coupling strength between the emitter 1 (2) with mode $ k_i $. 
	
	(c) Assuming that the giant atom (waveguide) is initially in the excited (vacuum) state, i.e., $ |\psi \left( t=0 \right) \rangle = |e,0\rangle $, we employ Qutip package ~\cite{Johansson12qutip,Johansson13qutip} to numerically solve the time-dependent Schr$ \ddot{\mathrm{o}} $dinger equation, which allows us to obtain the probability of the emitter $ |c_e(t)|^2 $ and the spontaneous decay rate. 
	
	Based on the method above, we plot all the dynamical evolutions in our work.
%%%%%%%%%%%%%%%%%%%%%%% References %%%%%%%%%%%%%%%%%%%%%%%%%

%%%%%%%%%% If using BibTeX:
%\bibliography{left-handed_ref}

\begin{thebibliography}{94}%
	\makeatletter
	\providecommand \@ifxundefined [1]{%
		\@ifx{#1\undefined}
	}%
	\providecommand \@ifnum [1]{%
		\ifnum #1\expandafter \@firstoftwo
		\else \expandafter \@secondoftwo
		\fi
	}%
	\providecommand \@ifx [1]{%
		\ifx #1\expandafter \@firstoftwo
		\else \expandafter \@secondoftwo
		\fi
	}%
	\providecommand \natexlab [1]{#1}%
	\providecommand \enquote  [1]{``#1''}%
	\providecommand \bibnamefont  [1]{#1}%
	\providecommand \bibfnamefont [1]{#1}%
	\providecommand \citenamefont [1]{#1}%
	\providecommand \href@noop [0]{\@secondoftwo}%
	\providecommand \href [0]{\begingroup \@sanitize@url \@href}%
	\providecommand \@href[1]{\@@startlink{#1}\@@href}%
	\providecommand \@@href[1]{\endgroup#1\@@endlink}%
	\providecommand \@sanitize@url [0]{\catcode `\\12\catcode `\$12\catcode
		`\&12\catcode `\#12\catcode `\^12\catcode `\_12\catcode `\%12\relax}%
	\providecommand \@@startlink[1]{}%
	\providecommand \@@endlink[0]{}%
	\providecommand \url  [0]{\begingroup\@sanitize@url \@url }%
	\providecommand \@url [1]{\endgroup\@href {#1}{\urlprefix }}%
	\providecommand \urlprefix  [0]{URL }%
	\providecommand \Eprint [0]{\href }%
	\providecommand \doibase [0]{https://doi.org/}%
	\providecommand \selectlanguage [0]{\@gobble}%
	\providecommand \bibinfo  [0]{\@secondoftwo}%
	\providecommand \bibfield  [0]{\@secondoftwo}%
	\providecommand \translation [1]{[#1]}%
	\providecommand \BibitemOpen [0]{}%
	\providecommand \bibitemStop [0]{}%
	\providecommand \bibitemNoStop [0]{.\EOS\space}%
	\providecommand \EOS [0]{\spacefactor3000\relax}%
	\providecommand \BibitemShut  [1]{\csname bibitem#1\endcsname}%
	\let\auto@bib@innerbib\@empty
	%</preamble>
	\bibitem [{\citenamefont {Sheremet}\ \emph {et~al.}(2023)\citenamefont
		{Sheremet}, \citenamefont {Petrov}, \citenamefont {Iorsh}, \citenamefont
		{Poshakinskiy},\ and\ \citenamefont {Poddubny}}]{RevModPhys.95.015002}%
	\BibitemOpen
	\bibfield  {author} {\bibinfo {author} {\bibfnamefont {A.~S.}\ \bibnamefont
			{Sheremet}}, \bibinfo {author} {\bibfnamefont {M.~I.}\ \bibnamefont
			{Petrov}}, \bibinfo {author} {\bibfnamefont {I.~V.}\ \bibnamefont {Iorsh}},
		\bibinfo {author} {\bibfnamefont {A.~V.}\ \bibnamefont {Poshakinskiy}},\ and\
		\bibinfo {author} {\bibfnamefont {A.~N.}\ \bibnamefont {Poddubny}},\
	}\bibfield  {title} {\bibinfo {title} {Waveguide quantum electrodynamics:
			{C}ollective radiance and photon-photon correlations},\ }\href
	{https://doi.org/10.1103/RevModPhys.95.015002} {\bibfield  {journal}
		{\bibinfo  {journal} {Rev. Mod. Phys.}\ }\textbf {\bibinfo {volume} {95}},\
		\bibinfo {pages} {015002} (\bibinfo {year} {2023})}\BibitemShut {NoStop}%
	\bibitem [{\citenamefont {Kannan}\ \emph
		{et~al.}(2020{\natexlab{a}})\citenamefont {Kannan}, \citenamefont
		{Ruckriegel}, \citenamefont {Campbell}, \citenamefont {Frisk~Kockum},
		\citenamefont {Braumüller}, \citenamefont {Kim}, \citenamefont {Kjaergaard},
		\citenamefont {Krantz}, \citenamefont {Melville}, \citenamefont
		{Niedzielski}, \citenamefont {Vepsäläinen}, \citenamefont {Winik},
		\citenamefont {Yoder}, \citenamefont {Nori}, \citenamefont {Orlando},
		\citenamefont {Gustavsson},\ and\ \citenamefont {Oliver}}]{RN1}%
	\BibitemOpen
	\bibfield  {author} {\bibinfo {author} {\bibfnamefont {B.}~\bibnamefont
			{Kannan}}, \bibinfo {author} {\bibfnamefont {M.~J.}\ \bibnamefont
			{Ruckriegel}}, \bibinfo {author} {\bibfnamefont {D.~L.}\ \bibnamefont
			{Campbell}}, \bibinfo {author} {\bibfnamefont {A.}~\bibnamefont
			{Frisk~Kockum}}, \bibinfo {author} {\bibfnamefont {J.}~\bibnamefont
			{Braumüller}}, \bibinfo {author} {\bibfnamefont {D.~K.}\ \bibnamefont
			{Kim}}, \bibinfo {author} {\bibfnamefont {M.}~\bibnamefont {Kjaergaard}},
		\bibinfo {author} {\bibfnamefont {P.}~\bibnamefont {Krantz}}, \bibinfo
		{author} {\bibfnamefont {A.}~\bibnamefont {Melville}}, \bibinfo {author}
		{\bibfnamefont {B.~M.}\ \bibnamefont {Niedzielski}}, \bibinfo {author}
		{\bibfnamefont {A.}~\bibnamefont {Vepsäläinen}}, \bibinfo {author}
		{\bibfnamefont {R.}~\bibnamefont {Winik}}, \bibinfo {author} {\bibfnamefont
			{J.~L.}\ \bibnamefont {Yoder}}, \bibinfo {author} {\bibfnamefont
			{F.}~\bibnamefont {Nori}}, \bibinfo {author} {\bibfnamefont {T.~P.}\
			\bibnamefont {Orlando}}, \bibinfo {author} {\bibfnamefont {S.}~\bibnamefont
			{Gustavsson}},\ and\ \bibinfo {author} {\bibfnamefont {W.~D.}\ \bibnamefont
			{Oliver}},\ }\bibfield  {title} {\bibinfo {title} {Waveguide quantum
			electrodynamics with superconducting artificial giant atoms},\ }\href
	{https://doi.org/10.1038/s41586-020-2529-9} {\bibfield  {journal} {\bibinfo
			{journal} {Nature}\ }\textbf {\bibinfo {volume} {583}},\ \bibinfo {pages}
		{775} (\bibinfo {year} {2020}{\natexlab{a}})}\BibitemShut {NoStop}%
	\bibitem [{\citenamefont {Cai}\ and\ \citenamefont
		{Jia}(2021)}]{PhysRevA.104.033710}%
	\BibitemOpen
	\bibfield  {author} {\bibinfo {author} {\bibfnamefont {Q.-Y.}\ \bibnamefont
			{Cai}}\ and\ \bibinfo {author} {\bibfnamefont {W.-Z.}\ \bibnamefont {Jia}},\
	}\bibfield  {title} {\bibinfo {title} {Coherent single-photon scattering
			spectra for a giant-atom waveguide-{QED} system beyond the dipole
			approximation},\ }\href {https://doi.org/10.1103/PhysRevA.104.033710}
	{\bibfield  {journal} {\bibinfo  {journal} {Phys. Rev. A}\ }\textbf {\bibinfo
			{volume} {104}},\ \bibinfo {pages} {033710} (\bibinfo {year}
		{2021})}\BibitemShut {NoStop}%
	\bibitem [{\citenamefont {Terradas-Brians\'o}\ \emph
		{et~al.}(2022)\citenamefont {Terradas-Brians\'o}, \citenamefont
		{Gonz\'alez-Guti\'errez}, \citenamefont {Nori}, \citenamefont
		{Mart\'{\i}n-Moreno},\ and\ \citenamefont {Zueco}}]{PhysRevA.106.063717}%
	\BibitemOpen
	\bibfield  {author} {\bibinfo {author} {\bibfnamefont {S.}~\bibnamefont
			{Terradas-Brians\'o}}, \bibinfo {author} {\bibfnamefont {C.~A.}\ \bibnamefont
			{Gonz\'alez-Guti\'errez}}, \bibinfo {author} {\bibfnamefont {F.}~\bibnamefont
			{Nori}}, \bibinfo {author} {\bibfnamefont {L.}~\bibnamefont
			{Mart\'{\i}n-Moreno}},\ and\ \bibinfo {author} {\bibfnamefont
			{D.}~\bibnamefont {Zueco}},\ }\bibfield  {title} {\bibinfo {title}
		{Ultrastrong waveguide {QED} with giant atoms},\ }\href
	{https://doi.org/10.1103/PhysRevA.106.063717} {\bibfield  {journal} {\bibinfo
			{journal} {Phys. Rev. A}\ }\textbf {\bibinfo {volume} {106}},\ \bibinfo
		{pages} {063717} (\bibinfo {year} {2022})}\BibitemShut {NoStop}%
	\bibitem [{\citenamefont {Kannan}\ \emph
		{et~al.}(2020{\natexlab{b}})\citenamefont {Kannan}, \citenamefont
		{Ruckriegel}, \citenamefont {Campbell}, \citenamefont {Frisk~Kockum},
		\citenamefont {Braumüller}, \citenamefont {Kim}, \citenamefont {Kjaergaard},
		\citenamefont {Krantz}, \citenamefont {Melville}, \citenamefont
		{Niedzielski}, \citenamefont {Vepsäläinen}, \citenamefont {Winik},
		\citenamefont {Yoder}, \citenamefont {Nori}, \citenamefont {Orlando},
		\citenamefont {Gustavsson},\ and\ \citenamefont {Oliver}}]{RN229}%
	\BibitemOpen
	\bibfield  {author} {\bibinfo {author} {\bibfnamefont {B.}~\bibnamefont
			{Kannan}}, \bibinfo {author} {\bibfnamefont {M.~J.}\ \bibnamefont
			{Ruckriegel}}, \bibinfo {author} {\bibfnamefont {D.~L.}\ \bibnamefont
			{Campbell}}, \bibinfo {author} {\bibfnamefont {A.}~\bibnamefont
			{Frisk~Kockum}}, \bibinfo {author} {\bibfnamefont {J.}~\bibnamefont
			{Braumüller}}, \bibinfo {author} {\bibfnamefont {D.~K.}\ \bibnamefont
			{Kim}}, \bibinfo {author} {\bibfnamefont {M.}~\bibnamefont {Kjaergaard}},
		\bibinfo {author} {\bibfnamefont {P.}~\bibnamefont {Krantz}}, \bibinfo
		{author} {\bibfnamefont {A.}~\bibnamefont {Melville}}, \bibinfo {author}
		{\bibfnamefont {B.~M.}\ \bibnamefont {Niedzielski}}, \bibinfo {author}
		{\bibfnamefont {A.}~\bibnamefont {Vepsäläinen}}, \bibinfo {author}
		{\bibfnamefont {R.}~\bibnamefont {Winik}}, \bibinfo {author} {\bibfnamefont
			{J.~L.}\ \bibnamefont {Yoder}}, \bibinfo {author} {\bibfnamefont
			{F.}~\bibnamefont {Nori}}, \bibinfo {author} {\bibfnamefont {T.~P.}\
			\bibnamefont {Orlando}}, \bibinfo {author} {\bibfnamefont {S.}~\bibnamefont
			{Gustavsson}},\ and\ \bibinfo {author} {\bibfnamefont {W.~D.}\ \bibnamefont
			{Oliver}},\ }\bibfield  {title} {\bibinfo {title} {Waveguide quantum
			electrodynamics with superconducting artificial giant atoms},\ }\href
	{https://doi.org/10.1038/s41586-020-2529-9} {\bibfield  {journal} {\bibinfo
			{journal} {Nature}\ }\textbf {\bibinfo {volume} {583}},\ \bibinfo {pages}
		{775} (\bibinfo {year} {2020}{\natexlab{b}})}\BibitemShut {NoStop}%
	\bibitem [{\citenamefont {Chen}\ \emph {et~al.}(2023)\citenamefont {Chen},
		\citenamefont {Du}, \citenamefont {Zhang}, \citenamefont {Guo}, \citenamefont
		{Wu}, \citenamefont {Artoni},\ and\ \citenamefont
		{Rocca}}]{chen2023giantatom}%
	\BibitemOpen
	\bibfield  {author} {\bibinfo {author} {\bibfnamefont {Y.-T.}\ \bibnamefont
			{Chen}}, \bibinfo {author} {\bibfnamefont {L.}~\bibnamefont {Du}}, \bibinfo
		{author} {\bibfnamefont {Y.}~\bibnamefont {Zhang}}, \bibinfo {author}
		{\bibfnamefont {L.}~\bibnamefont {Guo}}, \bibinfo {author} {\bibfnamefont
			{J.-H.}\ \bibnamefont {Wu}}, \bibinfo {author} {\bibfnamefont
			{M.}~\bibnamefont {Artoni}},\ and\ \bibinfo {author} {\bibfnamefont
			{G.~C.~L.}\ \bibnamefont {Rocca}},\ }\href@noop {} {\bibinfo {title}
		{Giant-atom {E}ffects on {P}opulation and {E}ntanglement {D}ynamics of
			{R}ydberg {A}toms}} (\bibinfo {year} {2023}),\ \Eprint
	{https://arxiv.org/abs/2304.14713} {arXiv:2304.14713 [quant-ph]} \BibitemShut
	{NoStop}%
	\bibitem [{\citenamefont {Yin}\ \emph {et~al.}(2022)\citenamefont {Yin},
		\citenamefont {Luo},\ and\ \citenamefont {Liao}}]{PhysRevA.106.063703}%
	\BibitemOpen
	\bibfield  {author} {\bibinfo {author} {\bibfnamefont {X.-L.}\ \bibnamefont
			{Yin}}, \bibinfo {author} {\bibfnamefont {W.-B.}\ \bibnamefont {Luo}},\ and\
		\bibinfo {author} {\bibfnamefont {J.-Q.}\ \bibnamefont {Liao}},\ }\bibfield
	{title} {\bibinfo {title} {Non-{M}arkovian disentanglement dynamics in
			double-giant-atom waveguide-{QED} systems},\ }\href
	{https://doi.org/10.1103/PhysRevA.106.063703} {\bibfield  {journal} {\bibinfo
			{journal} {Phys. Rev. A}\ }\textbf {\bibinfo {volume} {106}},\ \bibinfo
		{pages} {063703} (\bibinfo {year} {2022})}\BibitemShut {NoStop}%
	\bibitem [{\citenamefont {Zhang}\ \emph {et~al.}(2022)\citenamefont {Zhang},
		\citenamefont {Zhu}, \citenamefont {Chen}, \citenamefont {Peng},
		\citenamefont {Yin}, \citenamefont {Yang}, \citenamefont {Zhao},
		\citenamefont {Lu}, \citenamefont {Chai}, \citenamefont {Xiong},\ and\
		\citenamefont {Tan}}]{phy.2022.1054299}%
	\BibitemOpen
	\bibfield  {author} {\bibinfo {author} {\bibfnamefont {Y.-Q.}\ \bibnamefont
			{Zhang}}, \bibinfo {author} {\bibfnamefont {Z.-H.}\ \bibnamefont {Zhu}},
		\bibinfo {author} {\bibfnamefont {K.-K.}\ \bibnamefont {Chen}}, \bibinfo
		{author} {\bibfnamefont {Z.-H.}\ \bibnamefont {Peng}}, \bibinfo {author}
		{\bibfnamefont {W.-J.}\ \bibnamefont {Yin}}, \bibinfo {author} {\bibfnamefont
			{Y.}~\bibnamefont {Yang}}, \bibinfo {author} {\bibfnamefont {Y.-Q.}\
			\bibnamefont {Zhao}}, \bibinfo {author} {\bibfnamefont {Z.-Y.}\ \bibnamefont
			{Lu}}, \bibinfo {author} {\bibfnamefont {Y.-F.}\ \bibnamefont {Chai}},
		\bibinfo {author} {\bibfnamefont {Z.-Z.}\ \bibnamefont {Xiong}},\ and\
		\bibinfo {author} {\bibfnamefont {L.}~\bibnamefont {Tan}},\ }\bibfield
	{title} {\bibinfo {title} {Controllable single-photon routing between two
			waveguides by two giant two-level atoms},\ }\href
	{https://doi.org/10.3389/fphy.2022.1054299} {\bibfield  {journal} {\bibinfo
			{journal} {Front. Phys.}\ }\textbf {\bibinfo {volume} {10}} (\bibinfo {year}
		{2022})}\BibitemShut {NoStop}%
	\bibitem [{\citenamefont {Kannan}\ \emph
		{et~al.}(2020{\natexlab{c}})\citenamefont {Kannan}, \citenamefont
		{Ruckriegel}, \citenamefont {Campbell}, \citenamefont {Frisk~Kockum},
		\citenamefont {Braumüller}, \citenamefont {Kim}, \citenamefont {Kjaergaard},
		\citenamefont {Krantz}, \citenamefont {Melville}, \citenamefont
		{Niedzielski}, \citenamefont {Vepsäläinen}, \citenamefont {Winik},
		\citenamefont {Yoder}, \citenamefont {Nori}, \citenamefont {Orlando},
		\citenamefont {Gustavsson},\ and\ \citenamefont {Oliver}}]{RN232}%
	\BibitemOpen
	\bibfield  {author} {\bibinfo {author} {\bibfnamefont {B.}~\bibnamefont
			{Kannan}}, \bibinfo {author} {\bibfnamefont {M.~J.}\ \bibnamefont
			{Ruckriegel}}, \bibinfo {author} {\bibfnamefont {D.~L.}\ \bibnamefont
			{Campbell}}, \bibinfo {author} {\bibfnamefont {A.}~\bibnamefont
			{Frisk~Kockum}}, \bibinfo {author} {\bibfnamefont {J.}~\bibnamefont
			{Braumüller}}, \bibinfo {author} {\bibfnamefont {D.~K.}\ \bibnamefont
			{Kim}}, \bibinfo {author} {\bibfnamefont {M.}~\bibnamefont {Kjaergaard}},
		\bibinfo {author} {\bibfnamefont {P.}~\bibnamefont {Krantz}}, \bibinfo
		{author} {\bibfnamefont {A.}~\bibnamefont {Melville}}, \bibinfo {author}
		{\bibfnamefont {B.~M.}\ \bibnamefont {Niedzielski}}, \bibinfo {author}
		{\bibfnamefont {A.}~\bibnamefont {Vepsäläinen}}, \bibinfo {author}
		{\bibfnamefont {R.}~\bibnamefont {Winik}}, \bibinfo {author} {\bibfnamefont
			{J.~L.}\ \bibnamefont {Yoder}}, \bibinfo {author} {\bibfnamefont
			{F.}~\bibnamefont {Nori}}, \bibinfo {author} {\bibfnamefont {T.~P.}\
			\bibnamefont {Orlando}}, \bibinfo {author} {\bibfnamefont {S.}~\bibnamefont
			{Gustavsson}},\ and\ \bibinfo {author} {\bibfnamefont {W.~D.}\ \bibnamefont
			{Oliver}},\ }\bibfield  {title} {\bibinfo {title} {Waveguide quantum
			electrodynamics with superconducting artificial giant atoms},\ }\href
	{https://doi.org/10.1038/s41586-020-2529-9} {\bibfield  {journal} {\bibinfo
			{journal} {Nature}\ }\textbf {\bibinfo {volume} {583}},\ \bibinfo {pages}
		{775} (\bibinfo {year} {2020}{\natexlab{c}})}\BibitemShut {NoStop}%
	\bibitem [{\citenamefont {Du}\ \emph {et~al.}(2022{\natexlab{a}})\citenamefont
		{Du}, \citenamefont {Zhang}, \citenamefont {Wu}, \citenamefont {Kockum},\
		and\ \citenamefont {Li}}]{PhysRevLett.128.223602}%
	\BibitemOpen
	\bibfield  {author} {\bibinfo {author} {\bibfnamefont {L.}~\bibnamefont
			{Du}}, \bibinfo {author} {\bibfnamefont {Y.}~\bibnamefont {Zhang}}, \bibinfo
		{author} {\bibfnamefont {J.-H.}\ \bibnamefont {Wu}}, \bibinfo {author}
		{\bibfnamefont {A.~F.}\ \bibnamefont {Kockum}},\ and\ \bibinfo {author}
		{\bibfnamefont {Y.}~\bibnamefont {Li}},\ }\bibfield  {title} {\bibinfo
		{title} {Giant atoms in a {S}ynthetic {F}requency {D}imension},\ }\href
	{https://doi.org/10.1103/PhysRevLett.128.223602} {\bibfield  {journal}
		{\bibinfo  {journal} {Phys. Rev. Lett.}\ }\textbf {\bibinfo {volume} {128}},\
		\bibinfo {pages} {223602} (\bibinfo {year} {2022}{\natexlab{a}})}\BibitemShut
	{NoStop}%
	\bibitem [{\citenamefont {Yang}\ \emph {et~al.}(2021)\citenamefont {Yang},
		\citenamefont {Han}, \citenamefont {Zheng}, \citenamefont {Lan},\ and\
		\citenamefont {Yu}}]{Yang_2021}%
	\BibitemOpen
	\bibfield  {author} {\bibinfo {author} {\bibfnamefont {X.-P.}\ \bibnamefont
			{Yang}}, \bibinfo {author} {\bibfnamefont {Z.-K.}\ \bibnamefont {Han}},
		\bibinfo {author} {\bibfnamefont {W.}~\bibnamefont {Zheng}}, \bibinfo
		{author} {\bibfnamefont {D.}~\bibnamefont {Lan}},\ and\ \bibinfo {author}
		{\bibfnamefont {Y.}~\bibnamefont {Yu}},\ }\bibfield  {title} {\bibinfo
		{title} {The interference between a giant atom and an internal resonator},\
	}\href {https://doi.org/10.1088/1572-9494/ac1e06} {\bibfield  {journal}
		{\bibinfo  {journal} {Commun. Theor. Phys.}\ }\textbf {\bibinfo {volume}
			{73}},\ \bibinfo {pages} {115104} (\bibinfo {year} {2021})}\BibitemShut
	{NoStop}%
	\bibitem [{\citenamefont {Frisk~Kockum}\ \emph {et~al.}(2014)\citenamefont
		{Frisk~Kockum}, \citenamefont {Delsing},\ and\ \citenamefont
		{Johansson}}]{PhysRevA.90.013837}%
	\BibitemOpen
	\bibfield  {author} {\bibinfo {author} {\bibfnamefont {A.}~\bibnamefont
			{Frisk~Kockum}}, \bibinfo {author} {\bibfnamefont {P.}~\bibnamefont
			{Delsing}},\ and\ \bibinfo {author} {\bibfnamefont {G.}~\bibnamefont
			{Johansson}},\ }\bibfield  {title} {\bibinfo {title} {Designing
			frequency-dependent relaxation rates and lamb shifts for a giant artificial
			atom},\ }\href {https://doi.org/10.1103/PhysRevA.90.013837} {\bibfield
		{journal} {\bibinfo  {journal} {Phys. Rev. A}\ }\textbf {\bibinfo {volume}
			{90}},\ \bibinfo {pages} {013837} (\bibinfo {year} {2014})}\BibitemShut
	{NoStop}%
	\bibitem [{\citenamefont {Du}\ and\ \citenamefont
		{Li}(2021)}]{PhysRevA.104.023712}%
	\BibitemOpen
	\bibfield  {author} {\bibinfo {author} {\bibfnamefont {L.}~\bibnamefont
			{Du}}\ and\ \bibinfo {author} {\bibfnamefont {Y.}~\bibnamefont {Li}},\
	}\bibfield  {title} {\bibinfo {title} {Single-photon frequency conversion via
			a giant $\mathrm{\ensuremath{\Lambda}}$-type atom},\ }\href
	{https://doi.org/10.1103/PhysRevA.104.023712} {\bibfield  {journal} {\bibinfo
			{journal} {Phys. Rev. A}\ }\textbf {\bibinfo {volume} {104}},\ \bibinfo
		{pages} {023712} (\bibinfo {year} {2021})}\BibitemShut {NoStop}%
	\bibitem [{\citenamefont {Du}\ \emph {et~al.}(2022{\natexlab{b}})\citenamefont
		{Du}, \citenamefont {Zhang},\ and\ \citenamefont {Li}}]{RN234}%
	\BibitemOpen
	\bibfield  {author} {\bibinfo {author} {\bibfnamefont {L.}~\bibnamefont
			{Du}}, \bibinfo {author} {\bibfnamefont {Y.}~\bibnamefont {Zhang}},\ and\
		\bibinfo {author} {\bibfnamefont {Y.}~\bibnamefont {Li}},\ }\bibfield
	{title} {\bibinfo {title} {A giant atom with modulated transition
			frequency},\ }\href {https://doi.org/10.1007/s11467-022-1215-9} {\bibfield
		{journal} {\bibinfo  {journal} {Front. Phys.}\ }\textbf {\bibinfo {volume}
			{18}},\ \bibinfo {pages} {12301} (\bibinfo {year}
		{2022}{\natexlab{b}})}\BibitemShut {NoStop}%
	\bibitem [{\citenamefont {Du}\ \emph {et~al.}(2022{\natexlab{c}})\citenamefont
		{Du}, \citenamefont {Chen}, \citenamefont {Zhang},\ and\ \citenamefont
		{Li}}]{PhysRevResearch.4.023198}%
	\BibitemOpen
	\bibfield  {author} {\bibinfo {author} {\bibfnamefont {L.}~\bibnamefont
			{Du}}, \bibinfo {author} {\bibfnamefont {Y.-T.}\ \bibnamefont {Chen}},
		\bibinfo {author} {\bibfnamefont {Y.}~\bibnamefont {Zhang}},\ and\ \bibinfo
		{author} {\bibfnamefont {Y.}~\bibnamefont {Li}},\ }\bibfield  {title}
	{\bibinfo {title} {Giant atoms with time-dependent couplings},\ }\href
	{https://doi.org/10.1103/PhysRevResearch.4.023198} {\bibfield  {journal}
		{\bibinfo  {journal} {Phys. Rev. Res.}\ }\textbf {\bibinfo {volume} {4}},\
		\bibinfo {pages} {023198} (\bibinfo {year} {2022}{\natexlab{c}})}\BibitemShut
	{NoStop}%
	\bibitem [{\citenamefont {Kockum}\ \emph {et~al.}(2018)\citenamefont {Kockum},
		\citenamefont {Johansson},\ and\ \citenamefont
		{Nori}}]{PhysRevLett.120.140404}%
	\BibitemOpen
	\bibfield  {author} {\bibinfo {author} {\bibfnamefont {A.~F.}\ \bibnamefont
			{Kockum}}, \bibinfo {author} {\bibfnamefont {G.}~\bibnamefont {Johansson}},\
		and\ \bibinfo {author} {\bibfnamefont {F.}~\bibnamefont {Nori}},\ }\bibfield
	{title} {\bibinfo {title} {Decoherence-{F}ree {I}nteraction between {G}iant
			{A}toms in {W}aveguide {Q}uantum {E}lectrodynamics},\ }\href
	{https://doi.org/10.1103/PhysRevLett.120.140404} {\bibfield  {journal}
		{\bibinfo  {journal} {Phys. Rev. Lett.}\ }\textbf {\bibinfo {volume} {120}},\
		\bibinfo {pages} {140404} (\bibinfo {year} {2018})}\BibitemShut {NoStop}%
	\bibitem [{\citenamefont {Du}\ \emph {et~al.}(2023{\natexlab{a}})\citenamefont
		{Du}, \citenamefont {Guo},\ and\ \citenamefont {Li}}]{PhysRevA.107.023705}%
	\BibitemOpen
	\bibfield  {author} {\bibinfo {author} {\bibfnamefont {L.}~\bibnamefont
			{Du}}, \bibinfo {author} {\bibfnamefont {L.-Z.}\ \bibnamefont {Guo}},\ and\
		\bibinfo {author} {\bibfnamefont {Y.}~\bibnamefont {Li}},\ }\bibfield
	{title} {\bibinfo {title} {Complex decoherence-free interactions between
			giant atoms},\ }\href {https://doi.org/10.1103/PhysRevA.107.023705}
	{\bibfield  {journal} {\bibinfo  {journal} {Phys. Rev. A}\ }\textbf {\bibinfo
			{volume} {107}},\ \bibinfo {pages} {023705} (\bibinfo {year}
		{2023}{\natexlab{a}})}\BibitemShut {NoStop}%
	\bibitem [{\citenamefont {Carollo}\ \emph {et~al.}(2020)\citenamefont
		{Carollo}, \citenamefont {Cilluffo},\ and\ \citenamefont
		{Ciccarello}}]{PhysRevResearch.2.043184}%
	\BibitemOpen
	\bibfield  {author} {\bibinfo {author} {\bibfnamefont {A.}~\bibnamefont
			{Carollo}}, \bibinfo {author} {\bibfnamefont {D.}~\bibnamefont {Cilluffo}},\
		and\ \bibinfo {author} {\bibfnamefont {F.}~\bibnamefont {Ciccarello}},\
	}\bibfield  {title} {\bibinfo {title} {Mechanism of decoherence-free coupling
			between giant atoms},\ }\href
	{https://doi.org/10.1103/PhysRevResearch.2.043184} {\bibfield  {journal}
		{\bibinfo  {journal} {Phys. Rev. Res.}\ }\textbf {\bibinfo {volume} {2}},\
		\bibinfo {pages} {043184} (\bibinfo {year} {2020})}\BibitemShut {NoStop}%
	\bibitem [{\citenamefont {Soro}\ \emph {et~al.}(2023)\citenamefont {Soro},
		\citenamefont {Mu\~noz},\ and\ \citenamefont {Kockum}}]{PhysRevA.107.013710}%
	\BibitemOpen
	\bibfield  {author} {\bibinfo {author} {\bibfnamefont {A.}~\bibnamefont
			{Soro}}, \bibinfo {author} {\bibfnamefont {C.~S.}\ \bibnamefont {Mu\~noz}},\
		and\ \bibinfo {author} {\bibfnamefont {A.~F.}\ \bibnamefont {Kockum}},\
	}\bibfield  {title} {\bibinfo {title} {Interaction between giant atoms in a
			one-dimensional structured environment},\ }\href
	{https://doi.org/10.1103/PhysRevA.107.013710} {\bibfield  {journal} {\bibinfo
			{journal} {Phys. Rev. A}\ }\textbf {\bibinfo {volume} {107}},\ \bibinfo
		{pages} {013710} (\bibinfo {year} {2023})}\BibitemShut {NoStop}%
	\bibitem [{\citenamefont {Lim}\ \emph {et~al.}(2023)\citenamefont {Lim},
		\citenamefont {Mok},\ and\ \citenamefont {Kwek}}]{PhysRevA.107.023716}%
	\BibitemOpen
	\bibfield  {author} {\bibinfo {author} {\bibfnamefont {K.~H.}\ \bibnamefont
			{Lim}}, \bibinfo {author} {\bibfnamefont {W.-K.}\ \bibnamefont {Mok}},\ and\
		\bibinfo {author} {\bibfnamefont {L.-C.}\ \bibnamefont {Kwek}},\ }\bibfield
	{title} {\bibinfo {title} {Oscillating bound states in non-markovian photonic
			lattices},\ }\href {https://doi.org/10.1103/PhysRevA.107.023716} {\bibfield
		{journal} {\bibinfo  {journal} {Phys. Rev. A}\ }\textbf {\bibinfo {volume}
			{107}},\ \bibinfo {pages} {023716} (\bibinfo {year} {2023})}\BibitemShut
	{NoStop}%
	\bibitem [{\citenamefont {Xiao}\ \emph {et~al.}(2022)\citenamefont {Xiao},
		\citenamefont {Wang}, \citenamefont {Li}, \citenamefont {Chen},\ and\
		\citenamefont {Yuan}}]{Xiao_2022}%
	\BibitemOpen
	\bibfield  {author} {\bibinfo {author} {\bibfnamefont {H.}~\bibnamefont
			{Xiao}}, \bibinfo {author} {\bibfnamefont {L.}~\bibnamefont {Wang}}, \bibinfo
		{author} {\bibfnamefont {Z.-H.}\ \bibnamefont {Li}}, \bibinfo {author}
		{\bibfnamefont {X.}~\bibnamefont {Chen}},\ and\ \bibinfo {author}
		{\bibfnamefont {L.}~\bibnamefont {Yuan}},\ }\bibfield  {title} {\bibinfo
		{title} {Bound state in a giant atom-modulated resonators system},\ }\href
	{https://doi.org/10.1038/s41534-022-00591-7} {\bibfield  {journal} {\bibinfo
			{journal} {npj Quantum Inf}\ }\textbf {\bibinfo {volume} {8}},\ \bibinfo
		{pages} {80} (\bibinfo {year} {2022})}\BibitemShut {NoStop}%
	\bibitem [{\citenamefont {Wang}\ \emph {et~al.}(2021)\citenamefont {Wang},
		\citenamefont {Liu}, \citenamefont {Kockum}, \citenamefont {Li},\ and\
		\citenamefont {Nori}}]{PhysRevLett.126.043602}%
	\BibitemOpen
	\bibfield  {author} {\bibinfo {author} {\bibfnamefont {X.}~\bibnamefont
			{Wang}}, \bibinfo {author} {\bibfnamefont {T.}~\bibnamefont {Liu}}, \bibinfo
		{author} {\bibfnamefont {A.~F.}\ \bibnamefont {Kockum}}, \bibinfo {author}
		{\bibfnamefont {H.-R.}\ \bibnamefont {Li}},\ and\ \bibinfo {author}
		{\bibfnamefont {F.}~\bibnamefont {Nori}},\ }\bibfield  {title} {\bibinfo
		{title} {Tunable {C}hiral {B}ound {S}tates with {G}iant {A}toms},\ }\href
	{https://doi.org/10.1103/PhysRevLett.126.043602} {\bibfield  {journal}
		{\bibinfo  {journal} {Phys. Rev. Lett.}\ }\textbf {\bibinfo {volume} {126}},\
		\bibinfo {pages} {043602} (\bibinfo {year} {2021})}\BibitemShut {NoStop}%
	\bibitem [{\citenamefont {Guo}\ \emph {et~al.}(2020{\natexlab{a}})\citenamefont
		{Guo}, \citenamefont {Kockum}, \citenamefont {Marquardt},\ and\ \citenamefont
		{Johansson}}]{PhysRevResearch.2.043014}%
	\BibitemOpen
	\bibfield  {author} {\bibinfo {author} {\bibfnamefont {L.}~\bibnamefont
			{Guo}}, \bibinfo {author} {\bibfnamefont {A.~F.}\ \bibnamefont {Kockum}},
		\bibinfo {author} {\bibfnamefont {F.}~\bibnamefont {Marquardt}},\ and\
		\bibinfo {author} {\bibfnamefont {G.}~\bibnamefont {Johansson}},\ }\bibfield
	{title} {\bibinfo {title} {Oscillating bound states for a giant atom},\
	}\href {https://doi.org/10.1103/PhysRevResearch.2.043014} {\bibfield
		{journal} {\bibinfo  {journal} {Phys. Rev. Res.}\ }\textbf {\bibinfo {volume}
			{2}},\ \bibinfo {pages} {043014} (\bibinfo {year}
		{2020}{\natexlab{a}})}\BibitemShut {NoStop}%
	\bibitem [{\citenamefont {Guo}\ \emph {et~al.}(2020{\natexlab{b}})\citenamefont
		{Guo}, \citenamefont {Wang}, \citenamefont {Purdy},\ and\ \citenamefont
		{Taylor}}]{PhysRevA.102.033706}%
	\BibitemOpen
	\bibfield  {author} {\bibinfo {author} {\bibfnamefont {S.}~\bibnamefont
			{Guo}}, \bibinfo {author} {\bibfnamefont {Y.}~\bibnamefont {Wang}}, \bibinfo
		{author} {\bibfnamefont {T.}~\bibnamefont {Purdy}},\ and\ \bibinfo {author}
		{\bibfnamefont {J.}~\bibnamefont {Taylor}},\ }\bibfield  {title} {\bibinfo
		{title} {Beyond spontaneous emission: {G}iant atom bounded in the
			continuum},\ }\href {https://doi.org/10.1103/PhysRevA.102.033706} {\bibfield
		{journal} {\bibinfo  {journal} {Phys. Rev. A}\ }\textbf {\bibinfo {volume}
			{102}},\ \bibinfo {pages} {033706} (\bibinfo {year}
		{2020}{\natexlab{b}})}\BibitemShut {NoStop}%
	\bibitem [{\citenamefont {Zhao}\ and\ \citenamefont
		{Wang}(2020)}]{PhysRevA.101.053855}%
	\BibitemOpen
	\bibfield  {author} {\bibinfo {author} {\bibfnamefont {W.}~\bibnamefont
			{Zhao}}\ and\ \bibinfo {author} {\bibfnamefont {Z.}~\bibnamefont {Wang}},\
	}\bibfield  {title} {\bibinfo {title} {Single-photon scattering and bound
			states in an atom-waveguide system with two or multiple coupling points},\
	}\href {https://doi.org/10.1103/PhysRevA.101.053855} {\bibfield  {journal}
		{\bibinfo  {journal} {Phys. Rev. A}\ }\textbf {\bibinfo {volume} {101}},\
		\bibinfo {pages} {053855} (\bibinfo {year} {2020})}\BibitemShut {NoStop}%
	\bibitem [{\citenamefont {Jia}\ and\ \citenamefont
		{Yu}(2023)}]{jia2023atomphoton}%
	\BibitemOpen
	\bibfield  {author} {\bibinfo {author} {\bibfnamefont {W.-Z.}\ \bibnamefont
			{Jia}}\ and\ \bibinfo {author} {\bibfnamefont {M.-T.}\ \bibnamefont {Yu}},\
	}\href@noop {} {\bibinfo {title} {Atom-photon dressed states in a
			waveguide-{QED} system with multiple giant atoms coupled to a resonator-array
			waveguide}} (\bibinfo {year} {2023}),\ \Eprint
	{https://arxiv.org/abs/2304.02072} {arXiv:2304.02072 [quant-ph]} \BibitemShut
	{NoStop}%
	\bibitem [{\citenamefont {Wang}\ and\ \citenamefont {Li}(2022)}]{Wang_2022-1}%
	\BibitemOpen
	\bibfield  {author} {\bibinfo {author} {\bibfnamefont {X.}~\bibnamefont
			{Wang}}\ and\ \bibinfo {author} {\bibfnamefont {H.-R.}\ \bibnamefont {Li}},\
	}\bibfield  {title} {\bibinfo {title} {Chiral quantum network with giant
			atoms},\ }\href {https://doi.org/10.1088/2058-9565/ac6a04} {\bibfield
		{journal} {\bibinfo  {journal} {Quantum Sci. Technol.}\ }\textbf {\bibinfo
			{volume} {7}},\ \bibinfo {pages} {035007} (\bibinfo {year}
		{2022})}\BibitemShut {NoStop}%
	\bibitem [{\citenamefont {Soro}\ and\ \citenamefont
		{Kockum}(2022)}]{PhysRevA.105.023712}%
	\BibitemOpen
	\bibfield  {author} {\bibinfo {author} {\bibfnamefont {A.}~\bibnamefont
			{Soro}}\ and\ \bibinfo {author} {\bibfnamefont {A.~F.}\ \bibnamefont
			{Kockum}},\ }\bibfield  {title} {\bibinfo {title} {Chiral quantum optics with
			giant atoms},\ }\href {https://doi.org/10.1103/PhysRevA.105.023712}
	{\bibfield  {journal} {\bibinfo  {journal} {Phys. Rev. A}\ }\textbf {\bibinfo
			{volume} {105}},\ \bibinfo {pages} {023712} (\bibinfo {year}
		{2022})}\BibitemShut {NoStop}%
	\bibitem [{\citenamefont {Chen}\ \emph {et~al.}(2022)\citenamefont {Chen},
		\citenamefont {Du}, \citenamefont {Guo}, \citenamefont {Wang}, \citenamefont
		{Zhang}, \citenamefont {Li},\ and\ \citenamefont {Wu}}]{Chen_2022}%
	\BibitemOpen
	\bibfield  {author} {\bibinfo {author} {\bibfnamefont {Y.-T.}\ \bibnamefont
			{Chen}}, \bibinfo {author} {\bibfnamefont {L.}~\bibnamefont {Du}}, \bibinfo
		{author} {\bibfnamefont {L.}~\bibnamefont {Guo}}, \bibinfo {author}
		{\bibfnamefont {Z.}~\bibnamefont {Wang}}, \bibinfo {author} {\bibfnamefont
			{Y.}~\bibnamefont {Zhang}}, \bibinfo {author} {\bibfnamefont
			{Y.}~\bibnamefont {Li}},\ and\ \bibinfo {author} {\bibfnamefont {J.-H.}\
			\bibnamefont {Wu}},\ }\bibfield  {title} {\bibinfo {title} {Nonreciprocal and
			chiral single-photon scattering for giant atoms},\ }\href
	{https://doi.org/10.1038/s42005-022-00991-3} {\bibfield  {journal} {\bibinfo
			{journal} {Commun. Phys.}\ }\textbf {\bibinfo {volume} {5}},\ \bibinfo
		{pages} {215} (\bibinfo {year} {2022})}\BibitemShut {NoStop}%
	\bibitem [{\citenamefont {Du}\ \emph {et~al.}(2023{\natexlab{b}})\citenamefont
		{Du}, \citenamefont {Chen}, \citenamefont {Zhang}, \citenamefont {Li},\ and\
		\citenamefont {Wu}}]{du2023decay}%
	\BibitemOpen
	\bibfield  {author} {\bibinfo {author} {\bibfnamefont {L.}~\bibnamefont
			{Du}}, \bibinfo {author} {\bibfnamefont {Y.-T.}\ \bibnamefont {Chen}},
		\bibinfo {author} {\bibfnamefont {Y.}~\bibnamefont {Zhang}}, \bibinfo
		{author} {\bibfnamefont {Y.}~\bibnamefont {Li}},\ and\ \bibinfo {author}
		{\bibfnamefont {J.-H.}\ \bibnamefont {Wu}},\ }\href@noop {} {\bibinfo {title}
		{Decay dynamics of a giant atom in a structured bath with broken
			time-reversal symmetry}} (\bibinfo {year} {2023}{\natexlab{b}}),\ \Eprint
	{https://arxiv.org/abs/2212.04208} {arXiv:2212.04208 [quant-ph]} \BibitemShut
	{NoStop}%
	\bibitem [{\citenamefont {Joshi}\ \emph {et~al.}(2023)\citenamefont {Joshi},
		\citenamefont {Yang},\ and\ \citenamefont
		{Mirhosseini}}]{PhysRevX.13.021039}%
	\BibitemOpen
	\bibfield  {author} {\bibinfo {author} {\bibfnamefont {C.}~\bibnamefont
			{Joshi}}, \bibinfo {author} {\bibfnamefont {F.}~\bibnamefont {Yang}},\ and\
		\bibinfo {author} {\bibfnamefont {M.}~\bibnamefont {Mirhosseini}},\
	}\bibfield  {title} {\bibinfo {title} {Resonance {F}luorescence of a {C}hiral
			{A}rtificial {A}tom},\ }\href {https://doi.org/10.1103/PhysRevX.13.021039}
	{\bibfield  {journal} {\bibinfo  {journal} {Phys. Rev. X}\ }\textbf {\bibinfo
			{volume} {13}},\ \bibinfo {pages} {021039} (\bibinfo {year}
		{2023})}\BibitemShut {NoStop}%
	\bibitem [{\citenamefont {Vadiraj}\ \emph {et~al.}(2021)\citenamefont
		{Vadiraj}, \citenamefont {Ask}, \citenamefont {McConkey}, \citenamefont
		{Nsanzineza}, \citenamefont {Chang}, \citenamefont {Kockum},\ and\
		\citenamefont {Wilson}}]{PhysRevA.103.023710}%
	\BibitemOpen
	\bibfield  {author} {\bibinfo {author} {\bibfnamefont {A.~M.}\ \bibnamefont
			{Vadiraj}}, \bibinfo {author} {\bibfnamefont {A.}~\bibnamefont {Ask}},
		\bibinfo {author} {\bibfnamefont {T.~G.}\ \bibnamefont {McConkey}}, \bibinfo
		{author} {\bibfnamefont {I.}~\bibnamefont {Nsanzineza}}, \bibinfo {author}
		{\bibfnamefont {C.~W.~S.}\ \bibnamefont {Chang}}, \bibinfo {author}
		{\bibfnamefont {A.~F.}\ \bibnamefont {Kockum}},\ and\ \bibinfo {author}
		{\bibfnamefont {C.~M.}\ \bibnamefont {Wilson}},\ }\bibfield  {title}
	{\bibinfo {title} {Engineering the level structure of a giant artificial atom
			in waveguide quantum electrodynamics},\ }\href
	{https://doi.org/10.1103/PhysRevA.103.023710} {\bibfield  {journal} {\bibinfo
			{journal} {Phys. Rev. A}\ }\textbf {\bibinfo {volume} {103}},\ \bibinfo
		{pages} {023710} (\bibinfo {year} {2021})}\BibitemShut {NoStop}%
	\bibitem [{\citenamefont {Sun}\ \emph {et~al.}(2023)\citenamefont {Sun},
		\citenamefont {Liu}, \citenamefont {Chen},\ and\ \citenamefont
		{Li}}]{Sun_2023}%
	\BibitemOpen
	\bibfield  {author} {\bibinfo {author} {\bibfnamefont {X.-J.}\ \bibnamefont
			{Sun}}, \bibinfo {author} {\bibfnamefont {W.-X.}\ \bibnamefont {Liu}},
		\bibinfo {author} {\bibfnamefont {H.}~\bibnamefont {Chen}},\ and\ \bibinfo
		{author} {\bibfnamefont {H.-R.}\ \bibnamefont {Li}},\ }\bibfield  {title}
	{\bibinfo {title} {Tunable single-photon nonreciprocal scattering and
			targeted router in a giant atom-waveguide system with chiral couplings},\
	}\href {https://doi.org/10.1088/1572-9494/acb6ee} {\bibfield  {journal}
		{\bibinfo  {journal} {Commun. Theor. Phys.}\ }\textbf {\bibinfo {volume}
			{75}},\ \bibinfo {pages} {035103} (\bibinfo {year} {2023})}\BibitemShut
	{NoStop}%
	\bibitem [{\citenamefont {Liu}\ \emph {et~al.}(2023)\citenamefont {Liu},
		\citenamefont {Jin}, \citenamefont {Liu}, \citenamefont {Ming},\ and\
		\citenamefont {Yang}}]{RN124}%
	\BibitemOpen
	\bibfield  {author} {\bibinfo {author} {\bibfnamefont {J.-Y.}\ \bibnamefont
			{Liu}}, \bibinfo {author} {\bibfnamefont {J.-W.}\ \bibnamefont {Jin}},
		\bibinfo {author} {\bibfnamefont {H.-Y.}\ \bibnamefont {Liu}}, \bibinfo
		{author} {\bibfnamefont {Y.}~\bibnamefont {Ming}},\ and\ \bibinfo {author}
		{\bibfnamefont {R.-C.}\ \bibnamefont {Yang}},\ }\bibfield  {title} {\bibinfo
		{title} {Optical multi-fano-like phenomena with giant atom–waveguide
			systems},\ }\href {https://doi.org/10.1007/s11128-022-03816-y} {\bibfield
		{journal} {\bibinfo  {journal} {Quantum Inf Process}\ }\textbf {\bibinfo
			{volume} {22}},\ \bibinfo {pages} {74} (\bibinfo {year} {2023})}\BibitemShut
	{NoStop}%
	\bibitem [{\citenamefont {Zheng}\ \emph {et~al.}(2023)\citenamefont {Zheng},
		\citenamefont {Zhang}, \citenamefont {Wang}, \citenamefont {Han},\ and\
		\citenamefont {Wang}}]{Zheng_2023}%
	\BibitemOpen
	\bibfield  {author} {\bibinfo {author} {\bibfnamefont {C.-M.}\ \bibnamefont
			{Zheng}}, \bibinfo {author} {\bibfnamefont {W.}~\bibnamefont {Zhang}},
		\bibinfo {author} {\bibfnamefont {D.-Y.}\ \bibnamefont {Wang}}, \bibinfo
		{author} {\bibfnamefont {X.}~\bibnamefont {Han}},\ and\ \bibinfo {author}
		{\bibfnamefont {H.-F.}\ \bibnamefont {Wang}},\ }\bibfield  {title} {\bibinfo
		{title} {Simultaneously enhanced photon blockades in two microwave cavities
			via driving a giant atom},\ }\href {https://doi.org/10.1088/1367-2630/accd8c}
	{\bibfield  {journal} {\bibinfo  {journal} {New J. Phys.}\ }\textbf {\bibinfo
			{volume} {25}},\ \bibinfo {pages} {043030} (\bibinfo {year}
		{2023})}\BibitemShut {NoStop}%
	\bibitem [{\citenamefont {Yu}\ \emph {et~al.}(2021)\citenamefont {Yu},
		\citenamefont {Wang},\ and\ \citenamefont {Wu}}]{PhysRevA.104.013720}%
	\BibitemOpen
	\bibfield  {author} {\bibinfo {author} {\bibfnamefont {H.}~\bibnamefont
			{Yu}}, \bibinfo {author} {\bibfnamefont {Z.}~\bibnamefont {Wang}},\ and\
		\bibinfo {author} {\bibfnamefont {J.-H.}\ \bibnamefont {Wu}},\ }\bibfield
	{title} {\bibinfo {title} {Entanglement preparation and nonreciprocal
			excitation evolution in giant atoms by controllable dissipation and
			coupling},\ }\href {https://doi.org/10.1103/PhysRevA.104.013720} {\bibfield
		{journal} {\bibinfo  {journal} {Phys. Rev. A}\ }\textbf {\bibinfo {volume}
			{104}},\ \bibinfo {pages} {013720} (\bibinfo {year} {2021})}\BibitemShut
	{NoStop}%
	\bibitem [{\citenamefont {Andersson}\ \emph {et~al.}(2019)\citenamefont
		{Andersson}, \citenamefont {Suri}, \citenamefont {Guo}, \citenamefont
		{Aref},\ and\ \citenamefont {Delsing}}]{Andersson_2019}%
	\BibitemOpen
	\bibfield  {author} {\bibinfo {author} {\bibfnamefont {G.}~\bibnamefont
			{Andersson}}, \bibinfo {author} {\bibfnamefont {B.}~\bibnamefont {Suri}},
		\bibinfo {author} {\bibfnamefont {L.}~\bibnamefont {Guo}}, \bibinfo {author}
		{\bibfnamefont {T.}~\bibnamefont {Aref}},\ and\ \bibinfo {author}
		{\bibfnamefont {P.}~\bibnamefont {Delsing}},\ }\bibfield  {title} {\bibinfo
		{title} {Non-exponential decay of a giant artificial atom},\ }\href
	{https://doi.org/10.1038/s41567-019-0605-6} {\bibfield  {journal} {\bibinfo
			{journal} {Nat. Phys.}\ }\textbf {\bibinfo {volume} {15}},\ \bibinfo {pages}
		{1123} (\bibinfo {year} {2019})}\BibitemShut {NoStop}%
	\bibitem [{\citenamefont {Gu}\ \emph {et~al.}(2023)\citenamefont {Gu},
		\citenamefont {Huang}, \citenamefont {Yi}, \citenamefont {Chen},
		\citenamefont {Sun},\ and\ \citenamefont {Tan}}]{gu2023correlated}%
	\BibitemOpen
	\bibfield  {author} {\bibinfo {author} {\bibfnamefont {W.}~\bibnamefont
			{Gu}}, \bibinfo {author} {\bibfnamefont {H.}~\bibnamefont {Huang}}, \bibinfo
		{author} {\bibfnamefont {Z.}~\bibnamefont {Yi}}, \bibinfo {author}
		{\bibfnamefont {L.}~\bibnamefont {Chen}}, \bibinfo {author} {\bibfnamefont
			{L.}~\bibnamefont {Sun}},\ and\ \bibinfo {author} {\bibfnamefont
			{H.}~\bibnamefont {Tan}},\ }\href@noop {} {\bibinfo {title} {Correlated
			two-photon scattering in a 1{D} waveguide coupled to two- or three-level
			giant atoms}} (\bibinfo {year} {2023}),\ \Eprint
	{https://arxiv.org/abs/2306.13836} {arXiv:2306.13836 [quant-ph]} \BibitemShut
	{NoStop}%
	\bibitem [{\citenamefont {Garziano}\ \emph {et~al.}(2014)\citenamefont
		{Garziano}, \citenamefont {Stassi}, \citenamefont {Ridolfo}, \citenamefont
		{Di~Stefano},\ and\ \citenamefont {Savasta}}]{PhysRevA.90.043817}%
	\BibitemOpen
	\bibfield  {author} {\bibinfo {author} {\bibfnamefont {L.}~\bibnamefont
			{Garziano}}, \bibinfo {author} {\bibfnamefont {R.}~\bibnamefont {Stassi}},
		\bibinfo {author} {\bibfnamefont {A.}~\bibnamefont {Ridolfo}}, \bibinfo
		{author} {\bibfnamefont {O.}~\bibnamefont {Di~Stefano}},\ and\ \bibinfo
		{author} {\bibfnamefont {S.}~\bibnamefont {Savasta}},\ }\bibfield  {title}
	{\bibinfo {title} {Vacuum-induced symmetry breaking in a superconducting
			quantum circuit},\ }\href {https://doi.org/10.1103/PhysRevA.90.043817}
	{\bibfield  {journal} {\bibinfo  {journal} {Phys. Rev. A}\ }\textbf {\bibinfo
			{volume} {90}},\ \bibinfo {pages} {043817} (\bibinfo {year}
		{2014})}\BibitemShut {NoStop}%
	\bibitem [{\citenamefont {Clarke}\ and\ \citenamefont
		{Wilhelm}(2008{\natexlab{a}})}]{RN159}%
	\BibitemOpen
	\bibfield  {author} {\bibinfo {author} {\bibfnamefont {J.}~\bibnamefont
			{Clarke}}\ and\ \bibinfo {author} {\bibfnamefont {F.~K.}\ \bibnamefont
			{Wilhelm}},\ }\bibfield  {title} {\bibinfo {title} {Superconducting quantum
			bits},\ }\href {https://doi.org/10.1038/nature07128} {\bibfield  {journal}
		{\bibinfo  {journal} {Nature}\ }\textbf {\bibinfo {volume} {453}},\ \bibinfo
		{pages} {1031} (\bibinfo {year} {2008}{\natexlab{a}})}\BibitemShut {NoStop}%
	\bibitem [{\citenamefont {Wendin}\ and\ \citenamefont
		{Shumeiko}(2005)}]{wendin2005}%
	\BibitemOpen
	\bibfield  {author} {\bibinfo {author} {\bibfnamefont {G.}~\bibnamefont
			{Wendin}}\ and\ \bibinfo {author} {\bibfnamefont {V.~S.}\ \bibnamefont
			{Shumeiko}},\ }\href@noop {} {\bibinfo {title} {Superconducting {Q}uantum
			{C}ircuits, {Q}ubits and {C}omputing}} (\bibinfo {year} {2005}),\ \Eprint
	{https://arxiv.org/abs/cond-mat/0508729} {arXiv:cond-mat/0508729}
	\BibitemShut {NoStop}%
	\bibitem [{\citenamefont {Blais}\ \emph {et~al.}(2021)\citenamefont {Blais},
		\citenamefont {Grimsmo}, \citenamefont {Girvin},\ and\ \citenamefont
		{Wallraff}}]{RevModPhys.93.025005}%
	\BibitemOpen
	\bibfield  {author} {\bibinfo {author} {\bibfnamefont {A.}~\bibnamefont
			{Blais}}, \bibinfo {author} {\bibfnamefont {A.~L.}\ \bibnamefont {Grimsmo}},
		\bibinfo {author} {\bibfnamefont {S.~M.}\ \bibnamefont {Girvin}},\ and\
		\bibinfo {author} {\bibfnamefont {A.}~\bibnamefont {Wallraff}},\ }\bibfield
	{title} {\bibinfo {title} {Circuit quantum electrodynamics},\ }\href
	{https://doi.org/10.1103/RevModPhys.93.025005} {\bibfield  {journal}
		{\bibinfo  {journal} {Rev. Mod. Phys.}\ }\textbf {\bibinfo {volume} {93}},\
		\bibinfo {pages} {025005} (\bibinfo {year} {2021})}\BibitemShut {NoStop}%
	\bibitem [{\citenamefont {Clarke}\ and\ \citenamefont
		{Wilhelm}(2008{\natexlab{b}})}]{RN2}%
	\BibitemOpen
	\bibfield  {author} {\bibinfo {author} {\bibfnamefont {J.}~\bibnamefont
			{Clarke}}\ and\ \bibinfo {author} {\bibfnamefont {F.~K.}\ \bibnamefont
			{Wilhelm}},\ }\bibfield  {title} {\bibinfo {title} {Superconducting quantum
			bits},\ }\href {https://doi.org/10.1038/nature07128} {\bibfield  {journal}
		{\bibinfo  {journal} {Nature}\ }\textbf {\bibinfo {volume} {453}},\ \bibinfo
		{pages} {1031} (\bibinfo {year} {2008}{\natexlab{b}})}\BibitemShut {NoStop}%
	\bibitem [{\citenamefont {Blais}\ \emph {et~al.}(2020)\citenamefont {Blais},
		\citenamefont {Girvin},\ and\ \citenamefont {Oliver}}]{RN3}%
	\BibitemOpen
	\bibfield  {author} {\bibinfo {author} {\bibfnamefont {A.}~\bibnamefont
			{Blais}}, \bibinfo {author} {\bibfnamefont {S.~M.}\ \bibnamefont {Girvin}},\
		and\ \bibinfo {author} {\bibfnamefont {W.~D.}\ \bibnamefont {Oliver}},\
	}\bibfield  {title} {\bibinfo {title} {Quantum information processing and
			quantum optics with circuit quantum electrodynamics},\ }\href
	{https://doi.org/10.1038/s41567-020-0806-z} {\bibfield  {journal} {\bibinfo
			{journal} {Nat. Phys.}\ }\textbf {\bibinfo {volume} {16}},\ \bibinfo
		{pages} {247} (\bibinfo {year} {2020})}\BibitemShut {NoStop}%
	\bibitem [{\citenamefont {Chang}\ \emph {et~al.}(2022)\citenamefont {Chang},
		\citenamefont {Dubyna}, \citenamefont {Chien}, \citenamefont {Chen},
		\citenamefont {Wu},\ and\ \citenamefont {Kuo}}]{RN4}%
	\BibitemOpen
	\bibfield  {author} {\bibinfo {author} {\bibfnamefont {Y.-H.}\ \bibnamefont
			{Chang}}, \bibinfo {author} {\bibfnamefont {D.}~\bibnamefont {Dubyna}},
		\bibinfo {author} {\bibfnamefont {W.-C.}\ \bibnamefont {Chien}}, \bibinfo
		{author} {\bibfnamefont {C.-H.}\ \bibnamefont {Chen}}, \bibinfo {author}
		{\bibfnamefont {C.-S.}\ \bibnamefont {Wu}},\ and\ \bibinfo {author}
		{\bibfnamefont {W.}~\bibnamefont {Kuo}},\ }\bibfield  {title} {\bibinfo
		{title} {Circuit quantum electrodynamics with dressed states of a
			superconducting artificial atom},\ }\href
	{https://doi.org/10.1038/s41598-022-26828-1} {\bibfield  {journal} {\bibinfo
			{journal} {Sci Rep}\ }\textbf {\bibinfo {volume} {12}},\ \bibinfo {pages}
		{22308} (\bibinfo {year} {2022})}\BibitemShut {NoStop}%
	\bibitem [{\citenamefont {Wang}\ \emph {et~al.}(2019)\citenamefont {Wang},
		\citenamefont {Zhuravel}, \citenamefont {Indrajeet}, \citenamefont
		{Taketani}, \citenamefont {Hutchings}, \citenamefont {Hao}, \citenamefont
		{Rouxinol}, \citenamefont {Wilhelm}, \citenamefont {LaHaye}, \citenamefont
		{Ustinov},\ and\ \citenamefont {Plourde}}]{PhysRevApplied.11.054062}%
	\BibitemOpen
	\bibfield  {author} {\bibinfo {author} {\bibfnamefont {H.}~\bibnamefont
			{Wang}}, \bibinfo {author} {\bibfnamefont {A.}~\bibnamefont {Zhuravel}},
		\bibinfo {author} {\bibfnamefont {S.}~\bibnamefont {Indrajeet}}, \bibinfo
		{author} {\bibfnamefont {B.}~\bibnamefont {Taketani}}, \bibinfo {author}
		{\bibfnamefont {M.}~\bibnamefont {Hutchings}}, \bibinfo {author}
		{\bibfnamefont {Y.}~\bibnamefont {Hao}}, \bibinfo {author} {\bibfnamefont
			{F.}~\bibnamefont {Rouxinol}}, \bibinfo {author} {\bibfnamefont
			{F.}~\bibnamefont {Wilhelm}}, \bibinfo {author} {\bibfnamefont
			{M.}~\bibnamefont {LaHaye}}, \bibinfo {author} {\bibfnamefont
			{A.}~\bibnamefont {Ustinov}},\ and\ \bibinfo {author} {\bibfnamefont
			{B.}~\bibnamefont {Plourde}},\ }\bibfield  {title} {\bibinfo {title} {Mode
			{S}tructure in {S}uperconducting {M}etamaterial {T}ransmission-{L}ine
			{R}esonators},\ }\href {https://doi.org/10.1103/PhysRevApplied.11.054062}
	{\bibfield  {journal} {\bibinfo  {journal} {Phys. Rev. Appl.}\ }\textbf
		{\bibinfo {volume} {11}},\ \bibinfo {pages} {054062} (\bibinfo {year}
		{2019})}\BibitemShut {NoStop}%
	\bibitem [{\citenamefont {Ferreira}\ \emph {et~al.}(2021)\citenamefont
		{Ferreira}, \citenamefont {Banker}, \citenamefont {Sipahigil}, \citenamefont
		{Matheny}, \citenamefont {Keller}, \citenamefont {Kim}, \citenamefont
		{Mirhosseini},\ and\ \citenamefont {Painter}}]{PhysRevX.11.041043}%
	\BibitemOpen
	\bibfield  {author} {\bibinfo {author} {\bibfnamefont {V.~S.}\ \bibnamefont
			{Ferreira}}, \bibinfo {author} {\bibfnamefont {J.}~\bibnamefont {Banker}},
		\bibinfo {author} {\bibfnamefont {A.}~\bibnamefont {Sipahigil}}, \bibinfo
		{author} {\bibfnamefont {M.~H.}\ \bibnamefont {Matheny}}, \bibinfo {author}
		{\bibfnamefont {A.~J.}\ \bibnamefont {Keller}}, \bibinfo {author}
		{\bibfnamefont {E.}~\bibnamefont {Kim}}, \bibinfo {author} {\bibfnamefont
			{M.}~\bibnamefont {Mirhosseini}},\ and\ \bibinfo {author} {\bibfnamefont
			{O.}~\bibnamefont {Painter}},\ }\bibfield  {title} {\bibinfo {title}
		{Collapse and {R}evival of an {A}rtificial {A}tom {C}oupled to a {S}tructured
			{P}hotonic {R}eservoir},\ }\href {https://doi.org/10.1103/PhysRevX.11.041043}
	{\bibfield  {journal} {\bibinfo  {journal} {Phys. Rev. X}\ }\textbf {\bibinfo
			{volume} {11}},\ \bibinfo {pages} {041043} (\bibinfo {year}
		{2021})}\BibitemShut {NoStop}%
	\bibitem [{\citenamefont {Lapine}\ \emph {et~al.}(2014)\citenamefont {Lapine},
		\citenamefont {Shadrivov},\ and\ \citenamefont
		{Kivshar}}]{RevModPhys.86.1093}%
	\BibitemOpen
	\bibfield  {author} {\bibinfo {author} {\bibfnamefont {M.}~\bibnamefont
			{Lapine}}, \bibinfo {author} {\bibfnamefont {I.~V.}\ \bibnamefont
			{Shadrivov}},\ and\ \bibinfo {author} {\bibfnamefont {Y.~S.}\ \bibnamefont
			{Kivshar}},\ }\bibfield  {title} {\bibinfo {title} {Colloquium: {N}onlinear
			metamaterials},\ }\href {https://doi.org/10.1103/RevModPhys.86.1093}
	{\bibfield  {journal} {\bibinfo  {journal} {Rev. Mod. Phys.}\ }\textbf
		{\bibinfo {volume} {86}},\ \bibinfo {pages} {1093} (\bibinfo {year}
		{2014})}\BibitemShut {NoStop}%
	\bibitem [{\citenamefont {Smith}\ \emph {et~al.}(2005)\citenamefont {Smith},
		\citenamefont {Mock}, \citenamefont {Starr},\ and\ \citenamefont
		{Schurig}}]{PhysRevE.71.036609}%
	\BibitemOpen
	\bibfield  {author} {\bibinfo {author} {\bibfnamefont {D.~R.}\ \bibnamefont
			{Smith}}, \bibinfo {author} {\bibfnamefont {J.~J.}\ \bibnamefont {Mock}},
		\bibinfo {author} {\bibfnamefont {A.~F.}\ \bibnamefont {Starr}},\ and\
		\bibinfo {author} {\bibfnamefont {D.}~\bibnamefont {Schurig}},\ }\bibfield
	{title} {\bibinfo {title} {Gradient index metamaterials},\ }\href
	{https://doi.org/10.1103/PhysRevE.71.036609} {\bibfield  {journal} {\bibinfo
			{journal} {Phys. Rev. E}\ }\textbf {\bibinfo {volume} {71}},\ \bibinfo
		{pages} {036609} (\bibinfo {year} {2005})}\BibitemShut {NoStop}%
	\bibitem [{\citenamefont {Sugino}\ \emph {et~al.}(2022)\citenamefont {Sugino},
		\citenamefont {Alshaqaq},\ and\ \citenamefont
		{Erturk}}]{PhysRevB.106.174304}%
	\BibitemOpen
	\bibfield  {author} {\bibinfo {author} {\bibfnamefont {C.}~\bibnamefont
			{Sugino}}, \bibinfo {author} {\bibfnamefont {M.}~\bibnamefont {Alshaqaq}},\
		and\ \bibinfo {author} {\bibfnamefont {A.}~\bibnamefont {Erturk}},\
	}\bibfield  {title} {\bibinfo {title} {Spatially programmable wave
			compression and signal enhancement in a piezoelectric metamaterial
			waveguide},\ }\href {https://doi.org/10.1103/PhysRevB.106.174304} {\bibfield
		{journal} {\bibinfo  {journal} {Phys. Rev. B}\ }\textbf {\bibinfo {volume}
			{106}},\ \bibinfo {pages} {174304} (\bibinfo {year} {2022})}\BibitemShut
	{NoStop}%
	\bibitem [{\citenamefont {Fan}\ \emph {et~al.}(2022)\citenamefont {Fan},
		\citenamefont {Averitt},\ and\ \citenamefont {Padilla}}]{FanAverittPadilla}%
	\BibitemOpen
	\bibfield  {author} {\bibinfo {author} {\bibfnamefont {K.}~\bibnamefont
			{Fan}}, \bibinfo {author} {\bibfnamefont {R.~D.}\ \bibnamefont {Averitt}},\
		and\ \bibinfo {author} {\bibfnamefont {W.~J.}\ \bibnamefont {Padilla}},\
	}\bibfield  {title} {\bibinfo {title} {Active and tunable nanophotonic
			metamaterials},\ }\href {https://doi.org/doi:10.1515/nanoph-2022-0188}
	{\bibfield  {journal} {\bibinfo  {journal} {Nanophotonics}\ }\textbf
		{\bibinfo {volume} {11}},\ \bibinfo {pages} {3769} (\bibinfo {year}
		{2022})}\BibitemShut {NoStop}%
	\bibitem [{\citenamefont {Kukolj}\ and\ \citenamefont {\ifmmode \check{C}\else
			\v{C}\fi{}ubrovi\ifmmode~\acute{c}\else
			\'{c}\fi{}}(2019)}]{PhysRevA.100.053853}%
	\BibitemOpen
	\bibfield  {author} {\bibinfo {author} {\bibfnamefont {T.}~\bibnamefont
			{Kukolj}}\ and\ \bibinfo {author} {\bibfnamefont {M.}~\bibnamefont {\ifmmode
				\check{C}\else \v{C}\fi{}ubrovi\ifmmode~\acute{c}\else \'{c}\fi{}}},\
	}\bibfield  {title} {\bibinfo {title} {Spontaneous isotropy breaking for
			vortices in nonlinear left-handed metamaterials},\ }\href
	{https://doi.org/10.1103/PhysRevA.100.053853} {\bibfield  {journal} {\bibinfo
			{journal} {Phys. Rev. A}\ }\textbf {\bibinfo {volume} {100}},\ \bibinfo
		{pages} {053853} (\bibinfo {year} {2019})}\BibitemShut {NoStop}%
	\bibitem [{\citenamefont {Wang}\ \emph
		{et~al.}(2022{\natexlab{a}})\citenamefont {Wang}, \citenamefont {Lin},
		\citenamefont {Li}, \citenamefont {Liu},\ and\ \citenamefont
		{Li}}]{Wang_2022}%
	\BibitemOpen
	\bibfield  {author} {\bibinfo {author} {\bibfnamefont {X.}~\bibnamefont
			{Wang}}, \bibinfo {author} {\bibfnamefont {Y.-F.}\ \bibnamefont {Lin}},
		\bibinfo {author} {\bibfnamefont {J.-Q.}\ \bibnamefont {Li}}, \bibinfo
		{author} {\bibfnamefont {W.-X.}\ \bibnamefont {Liu}},\ and\ \bibinfo {author}
		{\bibfnamefont {H.-R.}\ \bibnamefont {Li}},\ }\bibfield  {title} {\bibinfo
		{title} {Chiral {SQUID}-metamaterial waveguide for circuit-{QED}},\ }\href
	{https://doi.org/10.1088/1367-2630/aca87e} {\bibfield  {journal} {\bibinfo
			{journal} {New J. Phys.}\ }\textbf {\bibinfo {volume} {24}},\ \bibinfo
		{pages} {123010} (\bibinfo {year} {2022}{\natexlab{a}})}\BibitemShut
	{NoStop}%
	\bibitem [{\citenamefont {Egger}\ and\ \citenamefont
		{Wilhelm}(2013)}]{PhysRevLett.111.163601}%
	\BibitemOpen
	\bibfield  {author} {\bibinfo {author} {\bibfnamefont {D.~J.}\ \bibnamefont
			{Egger}}\ and\ \bibinfo {author} {\bibfnamefont {F.~K.}\ \bibnamefont
			{Wilhelm}},\ }\bibfield  {title} {\bibinfo {title} {Multimode {C}ircuit
			{Q}uantum {E}lectrodynamics with {H}ybrid {M}etamaterial {T}ransmission
			{L}ines},\ }\href {https://doi.org/10.1103/PhysRevLett.111.163601} {\bibfield
		{journal} {\bibinfo  {journal} {Phys. Rev. Lett.}\ }\textbf {\bibinfo
			{volume} {111}},\ \bibinfo {pages} {163601} (\bibinfo {year}
		{2013})}\BibitemShut {NoStop}%
	\bibitem [{\citenamefont {Inoue}\ \emph {et~al.}(2023)\citenamefont {Inoue},
		\citenamefont {Noguchi}, \citenamefont {Yoshida}, \citenamefont {Kim},
		\citenamefont {Asano},\ and\ \citenamefont
		{Noda}}]{PhysRevApplied.20.L011001}%
	\BibitemOpen
	\bibfield  {author} {\bibinfo {author} {\bibfnamefont {T.}~\bibnamefont
			{Inoue}}, \bibinfo {author} {\bibfnamefont {N.}~\bibnamefont {Noguchi}},
		\bibinfo {author} {\bibfnamefont {M.}~\bibnamefont {Yoshida}}, \bibinfo
		{author} {\bibfnamefont {H.}~\bibnamefont {Kim}}, \bibinfo {author}
		{\bibfnamefont {T.}~\bibnamefont {Asano}},\ and\ \bibinfo {author}
		{\bibfnamefont {S.}~\bibnamefont {Noda}},\ }\bibfield  {title} {\bibinfo
		{title} {Unidirectional {P}erfect {R}eflection and {R}adiation in
			{D}ouble-{L}attice {P}hotonic {C}rystals},\ }\href
	{https://doi.org/10.1103/PhysRevApplied.20.L011001} {\bibfield  {journal}
		{\bibinfo  {journal} {Phys. Rev. Appl.}\ }\textbf {\bibinfo {volume} {20}},\
		\bibinfo {pages} {L011001} (\bibinfo {year} {2023})}\BibitemShut {NoStop}%
	\bibitem [{\citenamefont {Douglas}\ \emph {et~al.}(2016)\citenamefont
		{Douglas}, \citenamefont {Caneva},\ and\ \citenamefont
		{Chang}}]{PhysRevX.6.031017}%
	\BibitemOpen
	\bibfield  {author} {\bibinfo {author} {\bibfnamefont {J.~S.}\ \bibnamefont
			{Douglas}}, \bibinfo {author} {\bibfnamefont {T.}~\bibnamefont {Caneva}},\
		and\ \bibinfo {author} {\bibfnamefont {D.~E.}\ \bibnamefont {Chang}},\
	}\bibfield  {title} {\bibinfo {title} {Photon {M}olecules in {A}tomic {G}ases
			{T}rapped {N}ear {P}hotonic {C}rystal {W}aveguides},\ }\href
	{https://doi.org/10.1103/PhysRevX.6.031017} {\bibfield  {journal} {\bibinfo
			{journal} {Phys. Rev. X}\ }\textbf {\bibinfo {volume} {6}},\ \bibinfo {pages}
		{031017} (\bibinfo {year} {2016})}\BibitemShut {NoStop}%
	\bibitem [{\citenamefont {Lang}\ \emph {et~al.}(2018)\citenamefont {Lang},
		\citenamefont {Wang}, \citenamefont {Wang},\ and\ \citenamefont
		{Chong}}]{PhysRevB.98.094307}%
	\BibitemOpen
	\bibfield  {author} {\bibinfo {author} {\bibfnamefont {L.-J.}\ \bibnamefont
			{Lang}}, \bibinfo {author} {\bibfnamefont {Y.}~\bibnamefont {Wang}}, \bibinfo
		{author} {\bibfnamefont {H.}~\bibnamefont {Wang}},\ and\ \bibinfo {author}
		{\bibfnamefont {Y.~D.}\ \bibnamefont {Chong}},\ }\bibfield  {title} {\bibinfo
		{title} {Effects of non-{H}ermiticity on {S}u-{S}chrieffer-{H}eeger defect
			states},\ }\href {https://doi.org/10.1103/PhysRevB.98.094307} {\bibfield
		{journal} {\bibinfo  {journal} {Phys. Rev. B}\ }\textbf {\bibinfo {volume}
			{98}},\ \bibinfo {pages} {094307} (\bibinfo {year} {2018})}\BibitemShut
	{NoStop}%
	\bibitem [{\citenamefont {Obana}\ \emph {et~al.}(2019)\citenamefont {Obana},
		\citenamefont {Liu},\ and\ \citenamefont
		{Wakabayashi}}]{PhysRevB.100.075437}%
	\BibitemOpen
	\bibfield  {author} {\bibinfo {author} {\bibfnamefont {D.}~\bibnamefont
			{Obana}}, \bibinfo {author} {\bibfnamefont {F.}~\bibnamefont {Liu}},\ and\
		\bibinfo {author} {\bibfnamefont {K.}~\bibnamefont {Wakabayashi}},\
	}\bibfield  {title} {\bibinfo {title} {Topological edge states in the
			{S}u-{S}chrieffer-{H}eeger model},\ }\href
	{https://doi.org/10.1103/PhysRevB.100.075437} {\bibfield  {journal} {\bibinfo
			{journal} {Phys. Rev. B}\ }\textbf {\bibinfo {volume} {100}},\ \bibinfo
		{pages} {075437} (\bibinfo {year} {2019})}\BibitemShut {NoStop}%
	\bibitem [{\citenamefont {Jung}\ \emph {et~al.}(2014)\citenamefont {Jung},
		\citenamefont {Ustinov},\ and\ \citenamefont {Anlage}}]{Jung_2014}%
	\BibitemOpen
	\bibfield  {author} {\bibinfo {author} {\bibfnamefont {P.}~\bibnamefont
			{Jung}}, \bibinfo {author} {\bibfnamefont {A.~V.}\ \bibnamefont {Ustinov}},\
		and\ \bibinfo {author} {\bibfnamefont {S.~M.}\ \bibnamefont {Anlage}},\
	}\bibfield  {title} {\bibinfo {title} {Progress in superconducting
			metamaterials},\ }\href {https://doi.org/10.1088/0953-2048/27/7/073001}
	{\bibfield  {journal} {\bibinfo  {journal} {Supercond. Sci. Technol.}\
		}\textbf {\bibinfo {volume} {27}},\ \bibinfo {pages} {073001} (\bibinfo
		{year} {2014})}\BibitemShut {NoStop}%
	\bibitem [{\citenamefont {Messinger}\ \emph {et~al.}(2019)\citenamefont
		{Messinger}, \citenamefont {Taketani},\ and\ \citenamefont
		{Wilhelm}}]{PhysRevA.99.032325}%
	\BibitemOpen
	\bibfield  {author} {\bibinfo {author} {\bibfnamefont {A.}~\bibnamefont
			{Messinger}}, \bibinfo {author} {\bibfnamefont {B.~G.}\ \bibnamefont
			{Taketani}},\ and\ \bibinfo {author} {\bibfnamefont {F.~K.}\ \bibnamefont
			{Wilhelm}},\ }\bibfield  {title} {\bibinfo {title} {Left-handed superlattice
			metamaterials for circuit {QED}},\ }\href
	{https://doi.org/10.1103/PhysRevA.99.032325} {\bibfield  {journal} {\bibinfo
			{journal} {Phys. Rev. A}\ }\textbf {\bibinfo {volume} {99}},\ \bibinfo
		{pages} {032325} (\bibinfo {year} {2019})}\BibitemShut {NoStop}%
	\bibitem [{\citenamefont {Indrajeet}\ \emph {et~al.}(2020)\citenamefont
		{Indrajeet}, \citenamefont {Wang}, \citenamefont {Hutchings}, \citenamefont
		{Taketani}, \citenamefont {Wilhelm}, \citenamefont {LaHaye},\ and\
		\citenamefont {Plourde}}]{PhysRevApplied.14.064033}%
	\BibitemOpen
	\bibfield  {author} {\bibinfo {author} {\bibfnamefont {S.}~\bibnamefont
			{Indrajeet}}, \bibinfo {author} {\bibfnamefont {H.}~\bibnamefont {Wang}},
		\bibinfo {author} {\bibfnamefont {M.}~\bibnamefont {Hutchings}}, \bibinfo
		{author} {\bibfnamefont {B.}~\bibnamefont {Taketani}}, \bibinfo {author}
		{\bibfnamefont {F.~K.}\ \bibnamefont {Wilhelm}}, \bibinfo {author}
		{\bibfnamefont {M.}~\bibnamefont {LaHaye}},\ and\ \bibinfo {author}
		{\bibfnamefont {B.}~\bibnamefont {Plourde}},\ }\bibfield  {title} {\bibinfo
		{title} {Coupling a {S}uperconducting {Q}ubit to a {L}eft-{H}anded
			{M}etamaterial {R}esonator},\ }\href
	{https://doi.org/10.1103/PhysRevApplied.14.064033} {\bibfield  {journal}
		{\bibinfo  {journal} {Phys. Rev. Appl.}\ }\textbf {\bibinfo {volume} {14}},\
		\bibinfo {pages} {064033} (\bibinfo {year} {2020})}\BibitemShut {NoStop}%
	\bibitem [{\citenamefont {Wei}\ and\ \citenamefont {Zhao}(2020)}]{RN233}%
	\BibitemOpen
	\bibfield  {author} {\bibinfo {author} {\bibfnamefont {X.-J.}\ \bibnamefont
			{Wei}}\ and\ \bibinfo {author} {\bibfnamefont {S.-C.}\ \bibnamefont {Zhao}},\
	}\bibfield  {title} {\bibinfo {title} {Left-handedness in the
			balanced/unbalanced resonance conditions of a quantized composite right-left
			handed transmission line},\ }\href
	{https://doi.org/10.1140/epjb/e2020-10046-1} {\bibfield  {journal} {\bibinfo
			{journal} {Eur. Phys. J. B}\ }\textbf {\bibinfo {volume} {93}},\ \bibinfo
		{pages} {81} (\bibinfo {year} {2020})}\BibitemShut {NoStop}%
	\bibitem [{\citenamefont {Liberal}\ and\ \citenamefont
		{Ziolkowski}(2021)}]{10.1063/5.0044103}%
	\BibitemOpen
	\bibfield  {author} {\bibinfo {author} {\bibfnamefont {I.}~\bibnamefont
			{Liberal}}\ and\ \bibinfo {author} {\bibfnamefont {R.~W.}\ \bibnamefont
			{Ziolkowski}},\ }\bibfield  {title} {\bibinfo {title} {{Nonperturbative decay
				dynamics in metamaterial waveguides}},\ }\href
	{https://doi.org/10.1063/5.0044103} {\bibfield  {journal} {\bibinfo
			{journal} {Appl. Phys. Lett.}\ }\textbf {\bibinfo {volume} {118}},\ \bibinfo
		{pages} {111103} (\bibinfo {year} {2021})}\BibitemShut {NoStop}%
	\bibitem [{\citenamefont {Wang}\ and\ \citenamefont
		{Lancaster}(2006)}]{1687908}%
	\BibitemOpen
	\bibfield  {author} {\bibinfo {author} {\bibfnamefont {Y.}~\bibnamefont
			{Wang}}\ and\ \bibinfo {author} {\bibfnamefont {M.}~\bibnamefont
			{Lancaster}},\ }\bibfield  {title} {\bibinfo {title} {High-{T}emperature
			{S}uperconducting {C}oplanar {L}eft-handed {T}ransmission {L}ines and
			{R}esonators},\ }\href {https://doi.org/10.1109/TASC.2006.873992} {\bibfield
		{journal} {\bibinfo  {journal} {IEEE Trans. Appl.
				Supercond.}\ }\textbf {\bibinfo {volume} {16}},\ \bibinfo {pages}
		{1893} (\bibinfo {year} {2006})}\BibitemShut {NoStop}%
	\bibitem [{\citenamefont {Du}\ \emph {et~al.}(2006)\citenamefont {Du},
		\citenamefont {Chen},\ and\ \citenamefont {Li}}]{PhysRevB.74.113105}%
	\BibitemOpen
	\bibfield  {author} {\bibinfo {author} {\bibfnamefont {C.}~\bibnamefont
			{Du}}, \bibinfo {author} {\bibfnamefont {H.}~\bibnamefont {Chen}},\ and\
		\bibinfo {author} {\bibfnamefont {S.}~\bibnamefont {Li}},\ }\bibfield
	{title} {\bibinfo {title} {Quantum left-handed metamaterial from
			superconducting quantum-interference devices},\ }\href
	{https://doi.org/10.1103/PhysRevB.74.113105} {\bibfield  {journal} {\bibinfo
			{journal} {Phys. Rev. B}\ }\textbf {\bibinfo {volume} {74}},\ \bibinfo
		{pages} {113105} (\bibinfo {year} {2006})}\BibitemShut {NoStop}%
	\bibitem [{\citenamefont {Lobet}\ \emph {et~al.}(2020)\citenamefont {Lobet},
		\citenamefont {Liberal}, \citenamefont {Knall}, \citenamefont {Alam},
		\citenamefont {Reshef}, \citenamefont {Boyd}, \citenamefont {Engheta},\ and\
		\citenamefont {Mazur}}]{RN235}%
	\BibitemOpen
	\bibfield  {author} {\bibinfo {author} {\bibfnamefont {M.}~\bibnamefont
			{Lobet}}, \bibinfo {author} {\bibfnamefont {I.}~\bibnamefont {Liberal}},
		\bibinfo {author} {\bibfnamefont {E.~N.}\ \bibnamefont {Knall}}, \bibinfo
		{author} {\bibfnamefont {M.~Z.}\ \bibnamefont {Alam}}, \bibinfo {author}
		{\bibfnamefont {O.}~\bibnamefont {Reshef}}, \bibinfo {author} {\bibfnamefont
			{R.~W.}\ \bibnamefont {Boyd}}, \bibinfo {author} {\bibfnamefont
			{N.}~\bibnamefont {Engheta}},\ and\ \bibinfo {author} {\bibfnamefont
			{E.}~\bibnamefont {Mazur}},\ }\bibfield  {title} {\bibinfo {title}
		{Fundamental {R}adiative {P}rocesses in {N}ear-{Z}ero-{I}ndex {M}edia of
			{V}arious {D}imensionalities},\ }\href
	{https://doi.org/10.1021/acsphotonics.0c00782} {\bibfield  {journal}
		{\bibinfo  {journal} {ACS Photonics}\ }\textbf {\bibinfo {volume} {7}},\
		\bibinfo {pages} {1965} (\bibinfo {year} {2020})}\BibitemShut {NoStop}%
	\bibitem [{\citenamefont {Al\`u}\ and\ \citenamefont
		{Engheta}(2009)}]{PhysRevLett.103.043902}%
	\BibitemOpen
	\bibfield  {author} {\bibinfo {author} {\bibfnamefont {A.}~\bibnamefont
			{Al\`u}}\ and\ \bibinfo {author} {\bibfnamefont {N.}~\bibnamefont
			{Engheta}},\ }\bibfield  {title} {\bibinfo {title} {Boosting {M}olecular
			{F}luorescence with a {P}lasmonic {N}anolauncher},\ }\href
	{https://doi.org/10.1103/PhysRevLett.103.043902} {\bibfield  {journal}
		{\bibinfo  {journal} {Phys. Rev. Lett.}\ }\textbf {\bibinfo {volume} {103}},\
		\bibinfo {pages} {043902} (\bibinfo {year} {2009})}\BibitemShut {NoStop}%
	\bibitem [{\citenamefont {Mahmoud}\ and\ \citenamefont
		{Engheta}(2014)}]{RN236}%
	\BibitemOpen
	\bibfield  {author} {\bibinfo {author} {\bibfnamefont {A.~M.}\ \bibnamefont
			{Mahmoud}}\ and\ \bibinfo {author} {\bibfnamefont {N.}~\bibnamefont
			{Engheta}},\ }\bibfield  {title} {\bibinfo {title} {Wave–matter
			interactions in epsilon-and-mu-near-zero structures},\ }\href
	{https://doi.org/10.1038/ncomms6638} {\bibfield  {journal} {\bibinfo
			{journal} {Nat. Commun}\ }\textbf {\bibinfo {volume} {5}},\ \bibinfo {pages}
		{5638} (\bibinfo {year} {2014})}\BibitemShut {NoStop}%
	\bibitem [{\citenamefont {Blais}\ \emph {et~al.}(2004)\citenamefont {Blais},
		\citenamefont {Huang}, \citenamefont {Wallraff}, \citenamefont {Girvin},\
		and\ \citenamefont {Schoelkopf}}]{PhysRevA.69.062320}%
	\BibitemOpen
	\bibfield  {author} {\bibinfo {author} {\bibfnamefont {A.}~\bibnamefont
			{Blais}}, \bibinfo {author} {\bibfnamefont {R.-S.}\ \bibnamefont {Huang}},
		\bibinfo {author} {\bibfnamefont {A.}~\bibnamefont {Wallraff}}, \bibinfo
		{author} {\bibfnamefont {S.~M.}\ \bibnamefont {Girvin}},\ and\ \bibinfo
		{author} {\bibfnamefont {R.~J.}\ \bibnamefont {Schoelkopf}},\ }\bibfield
	{title} {\bibinfo {title} {Cavity quantum electrodynamics for superconducting
			electrical circuits: {A}n architecture for quantum computation},\ }\href
	{https://doi.org/10.1103/PhysRevA.69.062320} {\bibfield  {journal} {\bibinfo
			{journal} {Phys. Rev. A}\ }\textbf {\bibinfo {volume} {69}},\ \bibinfo
		{pages} {062320} (\bibinfo {year} {2004})}\BibitemShut {NoStop}%
	\bibitem [{\citenamefont {Koch}\ \emph {et~al.}(2007)\citenamefont {Koch},
		\citenamefont {Yu}, \citenamefont {Gambetta}, \citenamefont {Houck},
		\citenamefont {Schuster}, \citenamefont {Majer}, \citenamefont {Blais},
		\citenamefont {Devoret}, \citenamefont {Girvin},\ and\ \citenamefont
		{Schoelkopf}}]{PhysRevA.76.042319}%
	\BibitemOpen
	\bibfield  {author} {\bibinfo {author} {\bibfnamefont {J.}~\bibnamefont
			{Koch}}, \bibinfo {author} {\bibfnamefont {T.~M.}\ \bibnamefont {Yu}},
		\bibinfo {author} {\bibfnamefont {J.}~\bibnamefont {Gambetta}}, \bibinfo
		{author} {\bibfnamefont {A.~A.}\ \bibnamefont {Houck}}, \bibinfo {author}
		{\bibfnamefont {D.~I.}\ \bibnamefont {Schuster}}, \bibinfo {author}
		{\bibfnamefont {J.}~\bibnamefont {Majer}}, \bibinfo {author} {\bibfnamefont
			{A.}~\bibnamefont {Blais}}, \bibinfo {author} {\bibfnamefont {M.~H.}\
			\bibnamefont {Devoret}}, \bibinfo {author} {\bibfnamefont {S.~M.}\
			\bibnamefont {Girvin}},\ and\ \bibinfo {author} {\bibfnamefont {R.~J.}\
			\bibnamefont {Schoelkopf}},\ }\bibfield  {title} {\bibinfo {title}
		{Charge-insensitive qubit design derived from the {C}ooper pair box},\ }\href
	{https://doi.org/10.1103/PhysRevA.76.042319} {\bibfield  {journal} {\bibinfo
			{journal} {Phys. Rev. A}\ }\textbf {\bibinfo {volume} {76}},\ \bibinfo
		{pages} {042319} (\bibinfo {year} {2007})}\BibitemShut {NoStop}%
	\bibitem [{\citenamefont {Kim}\ \emph {et~al.}(2021)\citenamefont {Kim},
		\citenamefont {Zhang}, \citenamefont {Ferreira}, \citenamefont {Banker},
		\citenamefont {Iverson}, \citenamefont {Sipahigil}, \citenamefont {Bello},
		\citenamefont {Gonz\'alez-Tudela}, \citenamefont {Mirhosseini},\ and\
		\citenamefont {Painter}}]{PhysRevX.11.011015}%
	\BibitemOpen
	\bibfield  {author} {\bibinfo {author} {\bibfnamefont {E.}~\bibnamefont
			{Kim}}, \bibinfo {author} {\bibfnamefont {X.}~\bibnamefont {Zhang}}, \bibinfo
		{author} {\bibfnamefont {V.~S.}\ \bibnamefont {Ferreira}}, \bibinfo {author}
		{\bibfnamefont {J.}~\bibnamefont {Banker}}, \bibinfo {author} {\bibfnamefont
			{J.~K.}\ \bibnamefont {Iverson}}, \bibinfo {author} {\bibfnamefont
			{A.}~\bibnamefont {Sipahigil}}, \bibinfo {author} {\bibfnamefont
			{M.}~\bibnamefont {Bello}}, \bibinfo {author} {\bibfnamefont
			{A.}~\bibnamefont {Gonz\'alez-Tudela}}, \bibinfo {author} {\bibfnamefont
			{M.}~\bibnamefont {Mirhosseini}},\ and\ \bibinfo {author} {\bibfnamefont
			{O.}~\bibnamefont {Painter}},\ }\bibfield  {title} {\bibinfo {title} {Quantum
			{E}lectrodynamics in a {T}opological {W}aveguide},\ }\href
	{https://doi.org/10.1103/PhysRevX.11.011015} {\bibfield  {journal} {\bibinfo
			{journal} {Phys. Rev. X}\ }\textbf {\bibinfo {volume} {11}},\ \bibinfo
		{pages} {011015} (\bibinfo {year} {2021})}\BibitemShut {NoStop}%
	\bibitem [{\citenamefont {Calzona}\ and\ \citenamefont
		{Carrega}(2022)}]{Calzona_2023}%
	\BibitemOpen
	\bibfield  {author} {\bibinfo {author} {\bibfnamefont {A.}~\bibnamefont
			{Calzona}}\ and\ \bibinfo {author} {\bibfnamefont {M.}~\bibnamefont
			{Carrega}},\ }\bibfield  {title} {\bibinfo {title} {Multi-mode architectures
			for noise-resilient superconducting qubits},\ }\href
	{https://doi.org/10.1088/1361-6668/acaa64} {\bibfield  {journal} {\bibinfo
			{journal} {Supercond. Sci. Technol.}\ }\textbf {\bibinfo {volume} {36}},\
		\bibinfo {pages} {023001} (\bibinfo {year} {2022})}\BibitemShut {NoStop}%
	\bibitem [{\citenamefont {Vaaranta}\ \emph {et~al.}(2022)\citenamefont
		{Vaaranta}, \citenamefont {Cattaneo},\ and\ \citenamefont
		{Lake}}]{PhysRevA.106.042605}%
	\BibitemOpen
	\bibfield  {author} {\bibinfo {author} {\bibfnamefont {A.}~\bibnamefont
			{Vaaranta}}, \bibinfo {author} {\bibfnamefont {M.}~\bibnamefont {Cattaneo}},\
		and\ \bibinfo {author} {\bibfnamefont {R.~E.}\ \bibnamefont {Lake}},\
	}\bibfield  {title} {\bibinfo {title} {Dynamics of a dispersively coupled
			transmon qubit in the presence of a noise source embedded in the control
			line},\ }\href {https://doi.org/10.1103/PhysRevA.106.042605} {\bibfield
		{journal} {\bibinfo  {journal} {Phys. Rev. A}\ }\textbf {\bibinfo {volume}
			{106}},\ \bibinfo {pages} {042605} (\bibinfo {year} {2022})}\BibitemShut
	{NoStop}%
	\bibitem [{\citenamefont {Gu}\ \emph {et~al.}(2017)\citenamefont {Gu},
		\citenamefont {Kockum}, \citenamefont {Miranowicz}, \citenamefont {xi~Liu},\
		and\ \citenamefont {Nori}}]{GU20171}%
	\BibitemOpen
	\bibfield  {author} {\bibinfo {author} {\bibfnamefont {X.}~\bibnamefont
			{Gu}}, \bibinfo {author} {\bibfnamefont {A.~F.}\ \bibnamefont {Kockum}},
		\bibinfo {author} {\bibfnamefont {A.}~\bibnamefont {Miranowicz}}, \bibinfo
		{author} {\bibfnamefont {Y.}~\bibnamefont {xi~Liu}},\ and\ \bibinfo {author}
		{\bibfnamefont {F.}~\bibnamefont {Nori}},\ }\bibfield  {title} {\bibinfo
		{title} {Microwave photonics with superconducting quantum circuits},\ }\href
	{https://doi.org/https://doi.org/10.1016/j.physrep.2017.10.002} {\bibfield
		{journal} {\bibinfo  {journal} {Phys. Rep}\ }\textbf {\bibinfo {volume}
			{718-719}},\ \bibinfo {pages} {1} (\bibinfo {year} {2017})}\BibitemShut
	{NoStop}%
	\bibitem [{\citenamefont {Krantz}\ \emph {et~al.}(2019)\citenamefont {Krantz},
		\citenamefont {Kjaergaard}, \citenamefont {Yan}, \citenamefont {Orlando},
		\citenamefont {Gustavsson},\ and\ \citenamefont {Oliver}}]{Krantz2019}%
	\BibitemOpen
	\bibfield  {author} {\bibinfo {author} {\bibfnamefont {P.}~\bibnamefont
			{Krantz}}, \bibinfo {author} {\bibfnamefont {M.}~\bibnamefont {Kjaergaard}},
		\bibinfo {author} {\bibfnamefont {F.}~\bibnamefont {Yan}}, \bibinfo {author}
		{\bibfnamefont {T.~P.}\ \bibnamefont {Orlando}}, \bibinfo {author}
		{\bibfnamefont {S.}~\bibnamefont {Gustavsson}},\ and\ \bibinfo {author}
		{\bibfnamefont {W.~D.}\ \bibnamefont {Oliver}},\ }\bibfield  {title}
	{\bibinfo {title} {{A quantum engineer's guide to superconducting qubits}},\
	}\href {https://doi.org/10.1063/1.5089550} {\bibfield  {journal} {\bibinfo
			{journal} {Appl. Phys. Rev.}\ }\textbf {\bibinfo {volume} {6}},\
		\bibinfo {pages} {021318} (\bibinfo {year} {2019})}\BibitemShut {NoStop}%
	\bibitem [{\citenamefont {Wei\ss{}l}\ \emph {et~al.}(2015)\citenamefont
		{Wei\ss{}l}, \citenamefont {K\"ung}, \citenamefont {Dumur}, \citenamefont
		{Feofanov}, \citenamefont {Matei}, \citenamefont {Naud}, \citenamefont
		{Buisson}, \citenamefont {Hekking},\ and\ \citenamefont
		{Guichard}}]{PhysRevB.92.104508}%
	\BibitemOpen
	\bibfield  {author} {\bibinfo {author} {\bibfnamefont {T.}~\bibnamefont
			{Wei\ss{}l}}, \bibinfo {author} {\bibfnamefont {B.}~\bibnamefont {K\"ung}},
		\bibinfo {author} {\bibfnamefont {E.}~\bibnamefont {Dumur}}, \bibinfo
		{author} {\bibfnamefont {A.~K.}\ \bibnamefont {Feofanov}}, \bibinfo {author}
		{\bibfnamefont {I.}~\bibnamefont {Matei}}, \bibinfo {author} {\bibfnamefont
			{C.}~\bibnamefont {Naud}}, \bibinfo {author} {\bibfnamefont {O.}~\bibnamefont
			{Buisson}}, \bibinfo {author} {\bibfnamefont {F.~W.~J.}\ \bibnamefont
			{Hekking}},\ and\ \bibinfo {author} {\bibfnamefont {W.}~\bibnamefont
			{Guichard}},\ }\bibfield  {title} {\bibinfo {title} {Kerr coefficients of
			plasma resonances in {J}osephson junction chains},\ }\href
	{https://doi.org/10.1103/PhysRevB.92.104508} {\bibfield  {journal} {\bibinfo
			{journal} {Phys. Rev. B}\ }\textbf {\bibinfo {volume} {92}},\ \bibinfo
		{pages} {104508} (\bibinfo {year} {2015})}\BibitemShut {NoStop}%
	\bibitem [{\citenamefont {Lodahl}\ \emph {et~al.}(2015)\citenamefont {Lodahl},
		\citenamefont {Mahmoodian},\ and\ \citenamefont
		{Stobbe}}]{RevModPhys.87.347}%
	\BibitemOpen
	\bibfield  {author} {\bibinfo {author} {\bibfnamefont {P.}~\bibnamefont
			{Lodahl}}, \bibinfo {author} {\bibfnamefont {S.}~\bibnamefont {Mahmoodian}},\
		and\ \bibinfo {author} {\bibfnamefont {S.}~\bibnamefont {Stobbe}},\
	}\bibfield  {title} {\bibinfo {title} {Interfacing single photons and single
			quantum dots with photonic nanostructures},\ }\href
	{https://doi.org/10.1103/RevModPhys.87.347} {\bibfield  {journal} {\bibinfo
			{journal} {Rev. Mod. Phys.}\ }\textbf {\bibinfo {volume} {87}},\ \bibinfo
		{pages} {347} (\bibinfo {year} {2015})}\BibitemShut {NoStop}%
	\bibitem [{\citenamefont {Wang}\ \emph
		{et~al.}(2022{\natexlab{b}})\citenamefont {Wang}, \citenamefont {Gao},
		\citenamefont {Li}, \citenamefont {Zhu},\ and\ \citenamefont
		{Li}}]{PhysRevA.106.043703}%
	\BibitemOpen
	\bibfield  {author} {\bibinfo {author} {\bibfnamefont {X.}~\bibnamefont
			{Wang}}, \bibinfo {author} {\bibfnamefont {Z.-M.}\ \bibnamefont {Gao}},
		\bibinfo {author} {\bibfnamefont {J.-Q.}\ \bibnamefont {Li}}, \bibinfo
		{author} {\bibfnamefont {H.-B.}\ \bibnamefont {Zhu}},\ and\ \bibinfo {author}
		{\bibfnamefont {H.-R.}\ \bibnamefont {Li}},\ }\bibfield  {title} {\bibinfo
		{title} {Unconventional quantum electrodynamics with a {H}ofstadter-ladder
			waveguide},\ }\href {https://doi.org/10.1103/PhysRevA.106.043703} {\bibfield
		{journal} {\bibinfo  {journal} {Phys. Rev. A}\ }\textbf {\bibinfo {volume}
			{106}},\ \bibinfo {pages} {043703} (\bibinfo {year}
		{2022}{\natexlab{b}})}\BibitemShut {NoStop}%
	\bibitem [{\citenamefont {Bialynicki-Birula}\ and\ \citenamefont
		{Bialynicka-Birula}(1997)}]{PhysRevLett.78.2539}%
	\BibitemOpen
	\bibfield  {author} {\bibinfo {author} {\bibfnamefont {I.}~\bibnamefont
			{Bialynicki-Birula}}\ and\ \bibinfo {author} {\bibfnamefont {Z.}~\bibnamefont
			{Bialynicka-Birula}},\ }\bibfield  {title} {\bibinfo {title} {Rotational
			{F}requency {S}hift},\ }\href {https://doi.org/10.1103/PhysRevLett.78.2539}
	{\bibfield  {journal} {\bibinfo  {journal} {Phys. Rev. Lett.}\ }\textbf
		{\bibinfo {volume} {78}},\ \bibinfo {pages} {2539} (\bibinfo {year}
		{1997})}\BibitemShut {NoStop}%
	\bibitem [{\citenamefont {Debierre}\ \emph {et~al.}(2015)\citenamefont
		{Debierre}, \citenamefont {Goessens}, \citenamefont {Brainis},\ and\
		\citenamefont {Durt}}]{PhysRevA.92.023825}%
	\BibitemOpen
	\bibfield  {author} {\bibinfo {author} {\bibfnamefont {V.}~\bibnamefont
			{Debierre}}, \bibinfo {author} {\bibfnamefont {I.}~\bibnamefont {Goessens}},
		\bibinfo {author} {\bibfnamefont {E.}~\bibnamefont {Brainis}},\ and\ \bibinfo
		{author} {\bibfnamefont {T.}~\bibnamefont {Durt}},\ }\bibfield  {title}
	{\bibinfo {title} {Fermi's golden rule beyond the zeno regime},\ }\href
	{https://doi.org/10.1103/PhysRevA.92.023825} {\bibfield  {journal} {\bibinfo
			{journal} {Phys. Rev. A}\ }\textbf {\bibinfo {volume} {92}},\ \bibinfo
		{pages} {023825} (\bibinfo {year} {2015})}\BibitemShut {NoStop}%
	\bibitem [{\citenamefont {Ghafoor}(2014)}]{Ghafoor_2014}%
	\BibitemOpen
	\bibfield  {author} {\bibinfo {author} {\bibfnamefont {F.}~\bibnamefont
			{Ghafoor}},\ }\bibfield  {title} {\bibinfo {title} {Autler–townes multiplet
			spectroscopy},\ }\href {https://doi.org/10.1088/1054-660X/24/3/035702}
	{\bibfield  {journal} {\bibinfo  {journal} {Laser Phys.}\ }\textbf
		{\bibinfo {volume} {24}},\ \bibinfo {pages} {035702} (\bibinfo {year}
		{2014})}\BibitemShut {NoStop}%
	\bibitem [{\citenamefont {Glaetzle}\ \emph {et~al.}(2010)\citenamefont
		{Glaetzle}, \citenamefont {Hammerer}, \citenamefont {Daley}, \citenamefont
		{Blatt},\ and\ \citenamefont {Zoller}}]{GLAETZLE2010758}%
	\BibitemOpen
	\bibfield  {author} {\bibinfo {author} {\bibfnamefont {A.}~\bibnamefont
			{Glaetzle}}, \bibinfo {author} {\bibfnamefont {K.}~\bibnamefont {Hammerer}},
		\bibinfo {author} {\bibfnamefont {A.}~\bibnamefont {Daley}}, \bibinfo
		{author} {\bibfnamefont {R.}~\bibnamefont {Blatt}},\ and\ \bibinfo {author}
		{\bibfnamefont {P.}~\bibnamefont {Zoller}},\ }\bibfield  {title} {\bibinfo
		{title} {A single trapped atom in front of an oscillating mirror},\ }\href
	{https://doi.org/https://doi.org/10.1016/j.optcom.2009.10.063} {\bibfield
		{journal} {\bibinfo  {journal} {Opt. Commun.}\ }\textbf {\bibinfo
			{volume} {283}},\ \bibinfo {pages} {758} (\bibinfo {year}
		{2010})}\BibitemShut {NoStop}%
	\bibitem [{\citenamefont {Scully}\ and\ \citenamefont
		{Zubairy}(1997)}]{scully_zubairy_1997}%
	\BibitemOpen
	\bibfield  {author} {\bibinfo {author} {\bibfnamefont {M.~O.}\ \bibnamefont
			{Scully}}\ and\ \bibinfo {author} {\bibfnamefont {M.~S.}\ \bibnamefont
			{Zubairy}},\ }\href {https://doi.org/10.1017/CBO9780511813993} {\emph
		{\bibinfo {title} {Quantum Optics}}}\ (\bibinfo  {publisher} {Cambridge
		University Press},\ \bibinfo {year} {1997})\BibitemShut {NoStop}%
	\bibitem [{RN1(2023)}]{RN130}%
	\BibitemOpen
	\bibfield  {title} {\bibinfo {title} {Observing the dynamics of photon bound
			states using a single quantum dot},\ }\href
	{https://doi.org/10.1038/s41567-023-01998-5} {\bibfield  {journal} {\bibinfo
			{journal} {Nat. Phys.}\ }\textbf {\bibinfo {volume} {19}},\ \bibinfo {pages}
		{785} (\bibinfo {year} {2023})}\BibitemShut {NoStop}%
	\bibitem [{\citenamefont {Calaj\'o}\ \emph {et~al.}(2016)\citenamefont
		{Calaj\'o}, \citenamefont {Ciccarello}, \citenamefont {Chang},\ and\
		\citenamefont {Rabl}}]{PhysRevA.93.033833}%
	\BibitemOpen
	\bibfield  {author} {\bibinfo {author} {\bibfnamefont {G.}~\bibnamefont
			{Calaj\'o}}, \bibinfo {author} {\bibfnamefont {F.}~\bibnamefont
			{Ciccarello}}, \bibinfo {author} {\bibfnamefont {D.}~\bibnamefont {Chang}},\
		and\ \bibinfo {author} {\bibfnamefont {P.}~\bibnamefont {Rabl}},\ }\bibfield
	{title} {\bibinfo {title} {Atom-field dressed states in slow-light waveguide
			{QED}},\ }\href {https://doi.org/10.1103/PhysRevA.93.033833} {\bibfield
		{journal} {\bibinfo  {journal} {Phys. Rev. A}\ }\textbf {\bibinfo {volume}
			{93}},\ \bibinfo {pages} {033833} (\bibinfo {year} {2016})}\BibitemShut
	{NoStop}%
	\bibitem [{\citenamefont {Gonz\'alez-Tudela}\ and\ \citenamefont
		{Cirac}(2017)}]{PhysRevA.96.043811}%
	\BibitemOpen
	\bibfield  {author} {\bibinfo {author} {\bibfnamefont {A.}~\bibnamefont
			{Gonz\'alez-Tudela}}\ and\ \bibinfo {author} {\bibfnamefont {J.~I.}\
			\bibnamefont {Cirac}},\ }\bibfield  {title} {\bibinfo {title} {Markovian and
			non-{M}arkovian dynamics of quantum emitters coupled to two-dimensional
			structured reservoirs},\ }\href {https://doi.org/10.1103/PhysRevA.96.043811}
	{\bibfield  {journal} {\bibinfo  {journal} {Phys. Rev. A}\ }\textbf {\bibinfo
			{volume} {96}},\ \bibinfo {pages} {043811} (\bibinfo {year}
		{2017})}\BibitemShut {NoStop}%
	\bibitem [{\citenamefont {Bello}\ \emph {et~al.}(2019)\citenamefont {Bello},
		\citenamefont {Platero}, \citenamefont {Cirac},\ and\ \citenamefont
		{González-Tudela}}]{R100}%
	\BibitemOpen
	\bibfield  {author} {\bibinfo {author} {\bibfnamefont {M.}~\bibnamefont
			{Bello}}, \bibinfo {author} {\bibfnamefont {G.}~\bibnamefont {Platero}},
		\bibinfo {author} {\bibfnamefont {J.~I.}\ \bibnamefont {Cirac}},\ and\
		\bibinfo {author} {\bibfnamefont {A.}~\bibnamefont {González-Tudela}},\
	}\bibfield  {title} {\bibinfo {title} {Unconventional quantum optics in
			topological waveguide {QED}},\ }\href
	{https://doi.org/10.1126/sciadv.aaw0297} {\bibfield  {journal} {\bibinfo
			{journal} {Sci. Adv.}\ }\textbf {\bibinfo {volume} {5}},\ \bibinfo {pages}
		{eaaw0297} (\bibinfo {year} {2019})}\BibitemShut {NoStop}%
	\bibitem [{\citenamefont {González-Tudela}\ \emph {et~al.}(2015)\citenamefont
		{González-Tudela}, \citenamefont {Hung}, \citenamefont {Chang},
		\citenamefont {Cirac},\ and\ \citenamefont {Kimble}}]{RN231}%
	\BibitemOpen
	\bibfield  {author} {\bibinfo {author} {\bibfnamefont {A.}~\bibnamefont
			{González-Tudela}}, \bibinfo {author} {\bibfnamefont {C.~L.}\ \bibnamefont
			{Hung}}, \bibinfo {author} {\bibfnamefont {D.~E.}\ \bibnamefont {Chang}},
		\bibinfo {author} {\bibfnamefont {J.~I.}\ \bibnamefont {Cirac}},\ and\
		\bibinfo {author} {\bibfnamefont {H.~J.}\ \bibnamefont {Kimble}},\ }\bibfield
	{title} {\bibinfo {title} {Subwavelength vacuum lattices and atom–atom
			interactions in two-dimensional photonic crystals},\ }\href
	{https://doi.org/10.1038/nphoton.2015.54} {\bibfield  {journal} {\bibinfo
			{journal} {Nat. Photon.}\ }\textbf {\bibinfo {volume} {9}},\ \bibinfo
		{pages} {320} (\bibinfo {year} {2015})}\BibitemShut {NoStop}%
	\bibitem [{\citenamefont {Santos}\ and\ \citenamefont
		{Bachelard}(2023)}]{PhysRevLett.130.053601}%
	\BibitemOpen
	\bibfield  {author} {\bibinfo {author} {\bibfnamefont {A.~C.}\ \bibnamefont
			{Santos}}\ and\ \bibinfo {author} {\bibfnamefont {R.}~\bibnamefont
			{Bachelard}},\ }\bibfield  {title} {\bibinfo {title} {Generation of
			{M}aximally {E}ntangled {L}ong-{L}ived {S}tates with {G}iant {A}toms in a
			{W}aveguide},\ }\href {https://doi.org/10.1103/PhysRevLett.130.053601}
	{\bibfield  {journal} {\bibinfo  {journal} {Phys. Rev. Lett.}\ }\textbf
		{\bibinfo {volume} {130}},\ \bibinfo {pages} {053601} (\bibinfo {year}
		{2023})}\BibitemShut {NoStop}%
	\bibitem [{\citenamefont {Shahmoon}\ and\ \citenamefont
		{Kurizki}(2013)}]{PhysRevA.87.033831}%
	\BibitemOpen
	\bibfield  {author} {\bibinfo {author} {\bibfnamefont {E.}~\bibnamefont
			{Shahmoon}}\ and\ \bibinfo {author} {\bibfnamefont {G.}~\bibnamefont
			{Kurizki}},\ }\bibfield  {title} {\bibinfo {title} {Nonradiative interaction
			and entanglement between distant atoms},\ }\href
	{https://doi.org/10.1103/PhysRevA.87.033831} {\bibfield  {journal} {\bibinfo
			{journal} {Phys. Rev. A}\ }\textbf {\bibinfo {volume} {87}},\ \bibinfo
		{pages} {033831} (\bibinfo {year} {2013})}\BibitemShut {NoStop}%
	\bibitem [{\citenamefont {Huang}\ \emph {et~al.}(2012)\citenamefont {Huang},
		\citenamefont {Chen}, \citenamefont {Jin}, \citenamefont {Liu},\ and\
		\citenamefont {Wang}}]{PhysRevA.85.053827}%
	\BibitemOpen
	\bibfield  {author} {\bibinfo {author} {\bibfnamefont {Y.-G.}\ \bibnamefont
			{Huang}}, \bibinfo {author} {\bibfnamefont {G.}~\bibnamefont {Chen}},
		\bibinfo {author} {\bibfnamefont {C.-J.}\ \bibnamefont {Jin}}, \bibinfo
		{author} {\bibfnamefont {W.~M.}\ \bibnamefont {Liu}},\ and\ \bibinfo {author}
		{\bibfnamefont {X.-H.}\ \bibnamefont {Wang}},\ }\bibfield  {title} {\bibinfo
		{title} {Dipole-dipole interaction in a photonic crystal nanocavity},\ }\href
	{https://doi.org/10.1103/PhysRevA.85.053827} {\bibfield  {journal} {\bibinfo
			{journal} {Phys. Rev. A}\ }\textbf {\bibinfo {volume} {85}},\ \bibinfo
		{pages} {053827} (\bibinfo {year} {2012})}\BibitemShut {NoStop}%
	\bibitem [{\citenamefont {James}\ and\ \citenamefont
		{Jerke}(2007)}]{effective}%
	\BibitemOpen
	\bibfield  {author} {\bibinfo {author} {\bibfnamefont {D.~F.}\ \bibnamefont
			{James}}\ and\ \bibinfo {author} {\bibfnamefont {J.}~\bibnamefont {Jerke}},\
	}\bibfield  {title} {\bibinfo {title} {Effective {H}amiltonian theory and its
			applications in quantum information},\ }\href
	{https://doi.org/10.1139/p07-060} {\bibfield  {journal} {\bibinfo  {journal}
			{Can. J. Phys.}\ }\textbf {\bibinfo {volume} {85}},\ \bibinfo {pages} {625}
		(\bibinfo {year} {2007})}\BibitemShut {NoStop}%
	\bibitem [{\citenamefont {Johansson}\ \emph {et~al.}(2012)\citenamefont
		{Johansson}, \citenamefont {Nation},\ and\ \citenamefont
		{Nori}}]{Johansson12qutip}%
	\BibitemOpen
	\bibfield  {author} {\bibinfo {author} {\bibfnamefont {J.~R.}\ \bibnamefont
			{Johansson}}, \bibinfo {author} {\bibfnamefont {P.~D.}\ \bibnamefont
			{Nation}},\ and\ \bibinfo {author} {\bibfnamefont {F.}~\bibnamefont {Nori}},\
	}\bibfield  {title} {\bibinfo {title} {Qutip: {A}n open-source {P}ython
			framework for the dynamics of open quantum systems},\ }\href
	{http://www.sciencedirect.com/science/article/pii/S0010465512000835}
	{\bibfield  {journal} {\bibinfo  {journal} {Comput. Phys. Commun.}\ }\textbf
		{\bibinfo {volume} {183}},\ \bibinfo {pages} {1760} (\bibinfo {year}
		{2012})}\BibitemShut {NoStop}%
	\bibitem [{\citenamefont {Johansson}\ \emph {et~al.}(2013)\citenamefont
		{Johansson}, \citenamefont {Nation},\ and\ \citenamefont
		{Nori}}]{Johansson13qutip}%
	\BibitemOpen
	\bibfield  {author} {\bibinfo {author} {\bibfnamefont {J.~R.}\ \bibnamefont
			{Johansson}}, \bibinfo {author} {\bibfnamefont {P.~D.}\ \bibnamefont
			{Nation}},\ and\ \bibinfo {author} {\bibfnamefont {F.}~\bibnamefont {Nori}},\
	}\bibfield  {title} {\bibinfo {title} {Qutip 2: {A} {P}ython framework for
			the dynamics of open quantum systems},\ }\href
	{http://www.sciencedirect.com/science/article/pii/S0010465512003955}
	{\bibfield  {journal} {\bibinfo  {journal} {Comput. Phys. Commun.}\ }\textbf
		{\bibinfo {volume} {184}},\ \bibinfo {pages} {1234} (\bibinfo {year}
		{2013})}\BibitemShut {NoStop}%
\end{thebibliography}

%%%%%%%%%% If preparing manually:

\end{document}